\documentclass{article}

\usepackage{PRIMEarxiv}

\usepackage[utf8]{inputenc} % allow utf-8 input
\usepackage[T1]{fontenc}    % use 8-bit T1 fonts
\usepackage{hyperref}       % hyperlinks
\usepackage{url}            % simple URL typesetting
\usepackage{booktabs}       % professional-quality tables
\usepackage{amsfonts}       % blackboard math symbols
\usepackage{nicefrac}       % compact symbols for 1/2, etc.
\usepackage{microtype}      % microtypography
\usepackage{lipsum}
\usepackage{fancyhdr}       % header
\usepackage{graphicx}       % graphics
\graphicspath{{media/}}     % organize your images and other figures under media/ folder
\usepackage{multirow}
\graphicspath{{images/}} 
\usepackage{subfigure}
\usepackage[colorinlistoftodos]{todonotes}
\usepackage{amsmath,amssymb}
\newcommand*\emptycirc{\tikz\draw (0,0) circle (1.0ex);} 

\NewDocumentCommand{\halfcirc}{ O{} }{%
    \begin{tikzpicture}
        \fill[white] (0,0) circle (1.0ex); 
        \fill[black] (0,0) -- (270:1ex) arc (270:90:1ex) -- cycle; 
        \draw[black] (0,0) circle (1.0ex);
    \end{tikzpicture}
}
\newcommand*\fullcirc{\tikz\fill[fill=black] (0,0) circle (1.0ex);}

\title{Privacy-preserving in Blockchain-based Federated Learning Systems}

\author{
  Sameera K. M. \\
  Department of Computer Applications,\\
  Cochin University of Science\\
  and Technology, India \\
  \texttt{sameerakm@cusat.ac.in} \\
  \And
  Serena Nicolazzo$^{*}$ \\
  Department of Computer Science,\\
  University of Milan, Italy \\
  \texttt{serena.nicolazzo@unimi.it} \\
  $^{*}$\textbf{Corresponding author}
  \And
  Marco Arazzi\\
  Department of Electrical, Computer and\\Biomedical Engineering,\\ University of Pavia, Italy \\
  \texttt{marco.arazzi01@universitadipavia.it} \\
  \And
  Antonino Nocera \\
  Department of Electrical, Computer and\\ Biomedical Engineering,\\ University of Pavia, Italy \\
  \texttt{antonino.nocera@unipv.it} \\
  \And
  Rafidha Rehiman K. A. \\
  Department of Computer Applications,\\
  Cochin University of Science\\
  and Technology, India \\
  \texttt{rafidharehimanka@cusat.ac.in} \\
  \And
  Vinod P. \\
  Department of Mathematics, \\
  University of Padua, Italy\\
  \texttt{vinod.p@cusat.ac.in }\\
  \texttt{vinod.puthuvath@unipd.it}\\
  \And
  Mauro Conti\\
  Department of Mathematics, \\
  University of Padua,\\
  Italy\\
  \texttt{mauro.conti@unipd.it} \\
}

\begin{document}
\maketitle

\begin{abstract}
Federated Learning (FL) has recently arisen as a revolutionary approach to collaborative training Machine Learning models. According to this novel framework, multiple participants train a global model collaboratively, coordinating with a central aggregator without sharing their local data. As FL gains popularity in diverse domains, security, and privacy concerns arise due to the distributed nature of this solution. Therefore, integrating this strategy with Blockchain technology has been consolidated as a preferred choice to ensure the privacy and security of participants. 

This paper explores the research efforts carried out by the scientific community to define privacy solutions in scenarios adopting Blockchain-Enabled FL. It comprehensively summarizes the background related to FL and Blockchain, evaluates existing architectures for their integration, and the primary attacks and possible countermeasures to guarantee privacy in this setting. Finally, it reviews the main application scenarios where Blockchain-Enabled FL approaches have been proficiently applied.
This survey can help academia and industry practitioners understand which theories and techniques exist to improve the performance of FL through Blockchain to preserve privacy and which are the main challenges and future directions in this novel and still under-explored context. We believe this work provides a novel contribution respect to the previous surveys and is a valuable tool to explore the current landscape, understand perspectives, and pave the way for advancements or improvements in this amalgamation of Blockchain and Federated Learning.
\end{abstract}

% keywords can be removed

\section{Introduction}

Federated Learning (FL, hereafter) has undergone a significant surge in popularity in recent years. This novel strategy enables training Machine Learning models directly on user devices or at the edge without centralizing raw data. Because sensitive data stays on the user's device, the risk of exposing information to possible data breaches is lowered, and user privacy is preserved. Moreover, collaboration among workers provides access to a large amount of data, which enhances performance, making models more efficient and scalable. While FL offers several advantages, it also has limitations and challenges. The main FL characteristics that expose it to new threats are {\em(i)} system heterogeneity, {\em(ii)} the need for a trustworthy central authority for the coordination of the processing of locally trained models, {\em(iii)} vulnerability to data falsification and inference attack, {\em(iv)} the lack of incentive mechanism for the participating nodes, {\em(v)} communication security, and {\em(vi)} regulatory complaints\cite{qu2022Blockchain,zhu2023Blockchain}.

Because current implementations of the FL system do not provide proper mechanisms to address these challenges, researchers have recently begun to study approaches that leverage Blockchain \cite{nakamoto2008bitcoin}.
This new technology, derived from the decentralized cryptocurrency system, has attracted the interest of industry and academia for its countless potential. It relies on the possibility of performing authentic and traceable transactions without a trusted third party and ensuring secure data storage and tracking. Hence, integrating Blockchain and FL can empower this last paradigm, ensuring data privacy, trust, and model security in decentralized collaborative Learning environments.

This study comprehensively reviews Blockchain-enabled FL, focusing mainly on data privacy. Although several papers analyze different aspects of the Blockchain-enabled FL paradigm, a systematic review of existing works on privacy still needs to be included. Moreover, we provide a novel perspective examining the possible attacks menacing privacy in this scenario and all the current adopted countermeasures present in literature to guarantee privacy through a Blockchain-enabled FL system. Specifically, our main aim is to examine the existing literature on privacy attacks in Blockchain-enabled Federated Learning~(BCFL) systems. Moreover, we organize related papers according to the type of solution they implemented for privacy preservation, such as differential privacy, homomorphic encryption, secure multiparty computation, reward-driven approaches, hybrid privacy approaches, and cross-chained Federated Learning. Lastly, we delve into the practical applications of BCFL in cutting-edge scenarios, including healthcare, Industry 5.0, and the Internet of Vehicles.
This survey provides several contributions, namely:

\begin{itemize}
    \item it introduces a conceptual introduction to both FL and Blockchain technologies. Moreover, it deep dives into the description of existing architectures for Blockchain-enabled FL, describing how Blockchain can tackle the current challenges for FL, especially those related to privacy.
    \item it identifies the primary attacks menacing data privacy in Blockchain-enabled FL systems and the recently investigated countermeasures involving privacy methods, such as homomorphic encryption, differential privacy, secure multiparty computation methods, reputation approaches,  and solutions relying on cross-chain FL. 
    \item It describes how several practical application scenarios in various industries can benefit from integrating Blockchain and FL.
    \item It discusses and examines Blockchain-enabled FL systems' future directions and open research problems.
\end{itemize}

This survey offers fresh perspectives on the new paradigm of Blockchain-enabled FL focused on privacy-preserving. We expect the conducted analysis to be helpful to practitioners and researchers in categorizing the high number of studies dealing with privacy-preserving Blockchain-enabled FL approaches and in highlighting potentially promising directions to motivate future research work.

The structure of this paper is outlined as follows. Section \ref{sec:relatedWork} introduces related survey studies, while Section \ref{sec:methodology} delves into the methodology employed for conducting this survey.  In section \ref{sec:background}, we overview both FL and Blockchain main concepts.
In Section \ref{sec:FLandBlock}, we analyze the state-of-the-art regarding the integration of FL and Blockchain technology, describing the main architectures.
Section \ref{sec:attack} addresses the primary privacy threats within Blockchain-enabled FL. Sections \ref{sec:solution1} and \ref{sec:solution2} focus on the possible solutions to preserve privacy in such a domain. Section \ref{sec:application} is devoted to the analysis of several application scenarios that benefit from Blockchain-enabled FL (such as Healthcare, Industrial IoT (IIoT), and the Internet of Vehicles). In Section \ref{sec:future}, we explore various unresolved challenges and provide insights for potential areas of future research. Ultimately, Section \ref{sec:conclusion} encapsulates our concluding remarks on the survey.

\section{Comparison with other survey articles}
\label{sec:relatedWork}

Although several related surveys have been conducted to explore the integration of Blockchain and FL from different perspectives, most of them focus on different aspects, issues, or application domains related to this combination \cite{qu2022Blockchain,li2022Blockchain,ali2021integration,nguyen2021Federated,issa2023Blockchain}. 

For instance, Qu et al. \cite{qu2022Blockchain} consider three problems of Blockchain-enabled FL, namely decentralization, incentive mechanism, and membership selection. They focus on attack categorization and evaluate the performance of existing countermeasures.
The paper presented in \cite{li2022Blockchain} describes the structural designs of Blockchain-enabled FL, the deployed platforms, and possible industrial applications. Moreover, it analyses the aspects of Blockchain that allow an improvement of the FL system, such as the node incentive mechanisms.

The authors of \cite{qammar2023securing} conduct a literature review on the integration of Blockchain in FL, analyzing 41 research studies published between the years 2016 to June 2022. They focus on several aspects of the BCFL system: security and privacy, record and reward, verification, and accountability. 

In \cite{ali2021integration,issa2023Blockchain}, the authors focus on Blockchain-based FL approaches for IoT applications. Specifically, \cite{ali2021integration} presents the notion of Blockchain, its application to IoT, and the privacy issues and possible countermeasures. Then, they introduce the FL application in IoT systems, devise a taxonomy, and present privacy threats in FL. The combination of these two paradigms is only briefly presented through an IoT-based use case. 
The work proposed by Nguyen et al. \cite{nguyen2021Federated} instead focuses on the applications of BCFL in mobile-edge computing domains, analyzing some critical aspects of system design, including communication cost, resource allocation, incentive mechanism, as well as aspects related to security and the safeguarding of privacy.

Several works rely on custom taxonomies to categorize related literature on Blockchain-based FL \cite{huang2022Blockchain,li2022Blockchain2,zhu2023Blockchain}.
In particular, in \cite{huang2022Blockchain}, the authors propose a taxonomy to categorize Blockchain-based FL systems referring to three distinct layers: the Blockchain, the training, and the aggregation layers. They briefly review and summarize representative work according to this taxonomy.

Similarly, the study presented in \cite{li2022Blockchain2} analyses 41 research papers between
2018 and 2021 deal with Blockchain-based FL methodologies in smart environments, categorizing work in a custom taxonomy. In particular, FL methodologies are divided
into public FL and private FL environments. In public Blockchain mechanisms, they investigate only vertical FL approaches, whereas, for private Blockchain mechanisms, they evaluate both horizontal FL and Federated Transfer Learning approaches.

Zhu et al. \cite{zhu2023Blockchain} rely on a categorization of BCFL models in three classes: decoupled, coupled, and overlapped, according to how the FL and Blockchain functions are integrated. Then, they use these classes to compare the advantages and disadvantages of the e state-of-the-art solution they considered.

The authors of \cite{chhetri2023survey} conduct a brief review of existing literature on Blockchain-based FL that addresses privacy challenges. They describe only 18 papers published mainly from 2019 to 2022 but do not consider all the possible approaches to guarantee privacy in BCFL.

Table \ref{tab:survey} summarizes the main topics addressed by related surveys and makes a comparison with our contribution. As visible from the table, none of the existing contributions cover the topics presented in the survey or consider papers in the temporal span we analyzed. Table \ref{tab:proportio_paper} outlines the distribution of articles, specifying the number of articles considered in each category related to existing surveys and our work, along with the corresponding publication years.

\begin{table*}
\centering
\caption{Summary of related surveys and their significant contributions to Our Work, specifically focused on Privacy attacks and privacy preservation approaches in BCFL. }
\resizebox{\textwidth}{!}{
\renewcommand{\arraystretch}{1.4}

\begin{tabular}{|l|p{1.8cm}p{1.8cm}p{1.8cm}p{1.3cm}cclcccccc|}
\hline
\multirow{4}*{\textbf{Survey paper}} & Literature Timeline& Blockchain and FL Background &  Proposed General Architecture &Privacy Attack in BCFL&\multicolumn{7}{c}{\multirow{3}{*}{Privacy Preserving Approaches in BCFL}}& \multirow{4}*{Applications}&
\\
\cline{6-12}
 & &   & &  &$C_1$ & $C_2$ &$C_3$ & $C_4$& $C_5$& $C_6$&$C_7$&&
\\
\hline
Ali et al.\cite{ali2021integration} & 2019-2020 & \fullcirc&\emptycirc & \emptycirc&\emptycirc &\emptycirc &\emptycirc&\emptycirc&\emptycirc&\emptycirc&\emptycirc&\halfcirc&\\
\hline

Nguyen et al.\cite{nguyen2021Federated} & 2019-2020  & \fullcirc &\fullcirc& \emptycirc& \halfcirc& \emptycirc&\emptycirc&\emptycirc&\halfcirc&\emptycirc&\emptycirc&\halfcirc&\\
\hline
Huang et al.\cite{huang2022Blockchain} & 2019-2020  &\emptycirc &\fullcirc&\emptycirc& \halfcirc &\emptycirc&\emptycirc&\emptycirc&\halfcirc&\emptycirc&\emptycirc&\emptycirc&\\
\hline
Li et al.\cite{li2022Blockchain} & 2020-2021 &\fullcirc & \fullcirc& \emptycirc&\emptycirc & \emptycirc&\emptycirc&\emptycirc&\halfcirc&\emptycirc    &\emptycirc&\halfcirc&\\
\hline

Qu et al.\cite{qu2022Blockchain} & 2019-2021 &\fullcirc &\fullcirc&\emptycirc &\fullcirc & \emptycirc&\emptycirc&\emptycirc&\halfcirc&\emptycirc&\emptycirc&\emptycirc&\\
\hline
Issa et al.\cite{issa2023Blockchain} & 2019-2021&\fullcirc &\fullcirc & \emptycirc& \halfcirc& \emptycirc&\emptycirc&\emptycirc&\emptycirc&\emptycirc&\emptycirc&\emptycirc&\\
\hline
Zhu et al.\cite{zhu2023Blockchain} & 2019-2021&\fullcirc &\fullcirc &\emptycirc &\halfcirc& \emptycirc&\emptycirc&\emptycirc&\halfcirc&\emptycirc&\emptycirc&\halfcirc&\\
\hline
Chhetri et al.~\cite{chhetri2023survey} & 2019-2022 &  \halfcirc & \emptycirc &\emptycirc &\emptycirc &\halfcirc&\halfcirc&\halfcirc&\emptycirc&\emptycirc&\emptycirc&\emptycirc&\\
\hline

Qammar et al.~\cite{qammar2023securing} & 2019-2022 & \halfcirc & \fullcirc&\emptycirc & \fullcirc&\emptycirc&\emptycirc&\emptycirc&\halfcirc&\emptycirc&\emptycirc&\halfcirc&\\
\hline
% qammar2023securing
\textbf{Our Work} & \textbf{2018-2023}& \fullcirc& \fullcirc & \fullcirc & \fullcirc & \fullcirc& \fullcirc& \fullcirc& \fullcirc& \fullcirc& \fullcirc& \fullcirc&\\
\hline
\end{tabular}}
\\
\vspace{0.3cm}
\footnotesize $C_1:$ BCFL architectures for security and privacy protection, $C_2:$BCFL with differential privacy based approach, $C_3:$BCFL with homomorphic encryption based approach, $C_4:$BCFL with secure multiparty computation based approach, $C_5:$BCFL with reward-driven based approach, $C_6:$BCFL with hybrid privacy approach, $C_7:$Using cross-chain based approach.
\\
\emptycirc{} denotes that the corresponding aspect has not been discussed, \halfcirc{} indicates a partial discussion, and \fullcirc{} signifies a comprehensive exploration.
\label{tab:survey}
\end{table*}

\section{Methodology}
\label{sec:methodology}

\subsection{Research Approach}
Blockchain-enabled Federated Learning has surfaced as a groundbreaking paradigm in the rapidly evolving technology landscape, presenting the prospect of decentralized and collaborative machine Learning while safeguarding data privacy. This study intricately explores the critical aspect of privacy within this innovative framework, meticulously examining potential threats and presenting effective mitigating strategies. It systematically explores Blockchain fundamentals and Federated Learning along with its categorization. It delves into relevant literature on privacy attacks and protection methods in BCFL.
Moreover, it highlights the essential need for privacy preservation in BCFL-focused applications across domains. The study aims to identify areas of concern, and the paper thoroughly examines the open issues and limitations faced by Blockchain-enabled Federated Learning. Finally, it discusses the future direction in these areas, primarily focusing on enhancing BCFL's privacy.

\subsection{Search Strategy}

To acquire pertinent information concerning Federated Learning based on Blockchain, we devised a search plan in line with our research objectives. We started by thoroughly exploring Google Scholar and Web of Science. Then, we expanded our investigation to reputable academic repositories such as IEEE Xplore, ACM Digital Library, ScienceDirect, and SpringerLink. Our search spanned publications from 2018 to 2023 to ensure a thorough review of recent research. Additionally, we formulated search terms using specific phrases and keywords to cover different aspects of Blockchain-enabled Federated Learning.% during our initial exploration.
\begin{figure*}
    \centering
    \includegraphics[scale=0.478]{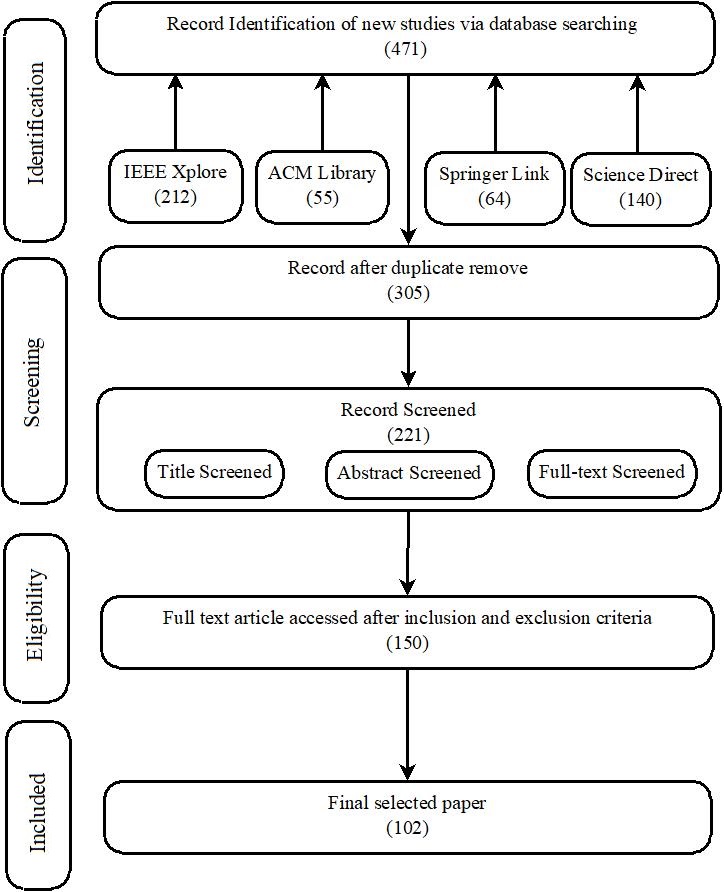}
    \caption{The PRISMA flow diagram visually outlines the various phases of the systematic review process.}
    \label{fig:prisma}
\end{figure*}
\par We consolidated the search terms using the conjunction operator (~AND) to pinpoint relevant studies accurately. Key search terms included \textquotedblleft Blockchain AND Federated Learning AND privacy\textquotedblright and \textquotedblleft privacy-preserving in Blockchain-based Federated Learning \textquotedblright. In addition to these primary terms, we incorporated supplementary search terms such as \textquotedblleft privacy attack \textquotedblright, \textquotedblleft inference attack\textquotedblright, \textquotedblleft homomorphic encryption\textquotedblright, \textquotedblleft differential privacy\textquotedblright, \textquotedblleft secure multiparty computation\textquotedblright, \textquotedblleft privacy-preserving in healthcare\textquotedblright, and \textquotedblleft internet of things\textquotedblright to enhance the comprehensiveness and scope of the search.
\par The PRISMA flow diagram in Figure~\ref{fig:prisma} visually illustrates the iterative screening process, depicting the counts of identified, excluded, and included research works.

%AN checkpoint

\subsection{Selection Criteria}

%In this section, we define the selection criteria used to determine whether a scientific work selected after applying our search criteria, is relevant and has enough quality to be included in this survey. A paper is selected if at least one inclusion criterion and none of the exclusion criteria apply.
This section outlines the criteria employed to assess the relevance and quality of scientific works selected for inclusion in this survey based on our search criteria. A paper is considered eligible for inclusion if it meets at least one of the following inclusion criteria and does not meet any exclusion criteria.

\subsubsection{Inclusion Criteria}
%To evaluate the relevance of a paper and include it in the present survey, we take the following criteria into account:
In assessing the relevance of a paper for inclusion in this survey, we consider the following criteria:

\begin{itemize}
    \item the importance of the corresponding author or supervisor in the domain under analysis;
    \item the citation count (primarily relying on Google Scholar \cite{Scholar} and Scopus \cite{Scopus} platforms);
    \item the age of the paper, we privileged more recent works;
    %\item the significance of the journal (or conference) where the paper has been published (we refer to Scimago \cite{Scimago} and Core.edu \cite{Core} as ranking websites for journals and conferences, respectively). 
     \item the paper's publication venue significance is assessed using Scimago \cite{Scimago} and Core.edu \cite{Core} rankings for journals and conferences, respectively.
\end{itemize}

\subsubsection{Exclusion Criteria}

Following the inclusion process, we apply the exclusion process, excluding a paper if it meets any of the following criteria:

\begin{itemize}
    \item the paper is not peer-reviewed;
    \item the paper is written in a language other than English;
    \item the date of publication exceeds six years w.r.t. our work (i.e., the paper publication year should be 2018 or later);
    \item the paper is not focused on solutions for Federated Learning-based Blockchain technology nor dealing with privacy;
    \item the paper lacks significance, as it represents an incremental improvement on a previously proposed approach, a duplicate publication, or an extended version of an already published key contribution.
\end{itemize}

%propotion of the paper considered in this survey 
\begin{table*}
\centering
\caption{Number of articles compared with existing surveys and Our Work, specifically focused on privacy attacks and privacy preservation in BCFL. }
\resizebox{\textwidth}{!}{
\renewcommand{\arraystretch}{1.4}

\begin{tabular}{|l|ccclcccccc|}
\hline
\multirow{2}*{\textbf{Survey paper}} &\multirow{2}*{Privacy Attack in BCFL} &\multicolumn{7}{c}{\multirow{1}{*}{Privacy Preserving Approaches in BCFL}}& \multirow{2}*{Applications}&
\\
\cline{3-9}
 &  &$C_1$ & $C_2$ &$C_3$ & $C_4$& $C_5$& $C_6$&$C_7$&&
\\
\hline
\multirow{2}*{Ali et al.\cite{ali2021integration}} & & & &&&&&&3&\\
 & & & &&&&&&(2018-2020)&\\
\hline

\multirow{2}*{Nguyen et al.\cite{nguyen2021Federated}} & & 6& &&&4&&&5&\\
 & & (2020-2021)& &&&(2019-2020)&&&(2018-2021)&\\
\hline
\multirow{2}*{Huang et al.\cite{huang2022Blockchain}} && 4 &&&&4&&&&\\
 &&(2021) &&&&(2020-2021)&&&&\\
\hline
\multirow{2}*{Li et al.\cite{li2022Blockchain} }& & & &&&1&    &&8&\\
& & & &&&(2019-2021)&    &&(2018-2021)&\\
\hline

\multirow{2}*{Qu et al.\cite{qu2022Blockchain} }&  &16 & &&&7 &&&&\\
&  &(2019-2021) & &&&(2019-2021)&&&&\\

\hline
\multirow{2}*{Issa et al.\cite{issa2023Blockchain}} &  &8 & &&&&&&&\\
 &  & (2019-2021)& &&&&&&&\\
\hline
\multirow{2}*{Zhu et al.\cite{zhu2023Blockchain} }&&5& &&&&&&2&\\
&&(2019-2020)& &&&&&&(2019-2021)&\\
\hline
\multirow{2}*{Chhetri et al.~\cite{chhetri2023survey}} & &&5&5&7&&&&&\\
& &&(2020-2022)&(2019-2021)&(2019-2021)&&&&&\\
\hline
\multirow{2}*{Qammar et al.~\cite{qammar2023securing} }&  & 4
&&&&1&&&2&\\
&  & 
(2020-2022)&&&&(2022)&&&(2019-2022)&\\
\hline
% qammar2023securing
\multirow{2}*{\textbf{Our Work}} &  2 & 14& 19& 14& 5&17& 9&4& 31&\\
 &  (2020-2023) & (2018-2023)& (2019-2023)& (2021-2023)& (2019-2023)&(2018-2023)& (2020-2022)&(2021-2023)& (2018-2023)&\\
\hline
\end{tabular}}
\\
\vspace{0.3cm}
\footnotesize $C_1:$ BCFL architectures for security and privacy protection, $C_2:$BCFL with differential privacy based approach, $C_3:$BCFL with homomorphic encryption based approach, $C_4:$BCFL with secure multiparty computation based approach, $C_5:$BCFL with reward-driven based approach, $C_6:$BCFL with hybrid privacy approach, $C_7:$Using cross-chain based approach.

\label{tab:proportio_paper}
\end{table*}

\section{Background Knowledge}
\label{sec:background}

This section provides the essential background information to contextualize our survey. In particular, we describe the main concepts related to FL, its workflow, and the principal categorizations and challenges of such an approach. Moreover, we illustrate the fundamental notions about Blockchain.
Table \ref{tab:SystemSymbols} summarizes the acronyms used in the paper.

\begin{table}

\centering
\caption{List of the acronyms used in the paper.\label{tab:SystemSymbols}}

\begin{tabular}{|l|l|}
\hline
    \textbf{Acronyms} & \textbf{Description}\\
    \hline \hline
    BCFL & Blockchain-enabled FL\\\hline
    DP & Differential Privacy\\\hline
    FL & Federated Learning\\\hline    
    HE & Homomorphic Encryption\\\hline
    IID & Independent and Identically Distributed\\ \hline
    IPFS & Interplanetary File System\\\hline
    ML & Machine Learning\\\hline
    SMPC & Secure Multiparty Computation\\\hline
    IoT &Internet of Things\\ \hline
    
\end{tabular}

\end{table}

\subsection{Centralized Learning, Distributed Learning, vs. Federated Learning} 

This part explains how ML architectures have evolved, progressing from centralized models to distributed on-site solutions and, most recently, up to Federated Learning (FL) \cite{abdulrahman2020survey}.

\begin{figure}[ht]
	\centerline{
        \includegraphics[scale=0.4]{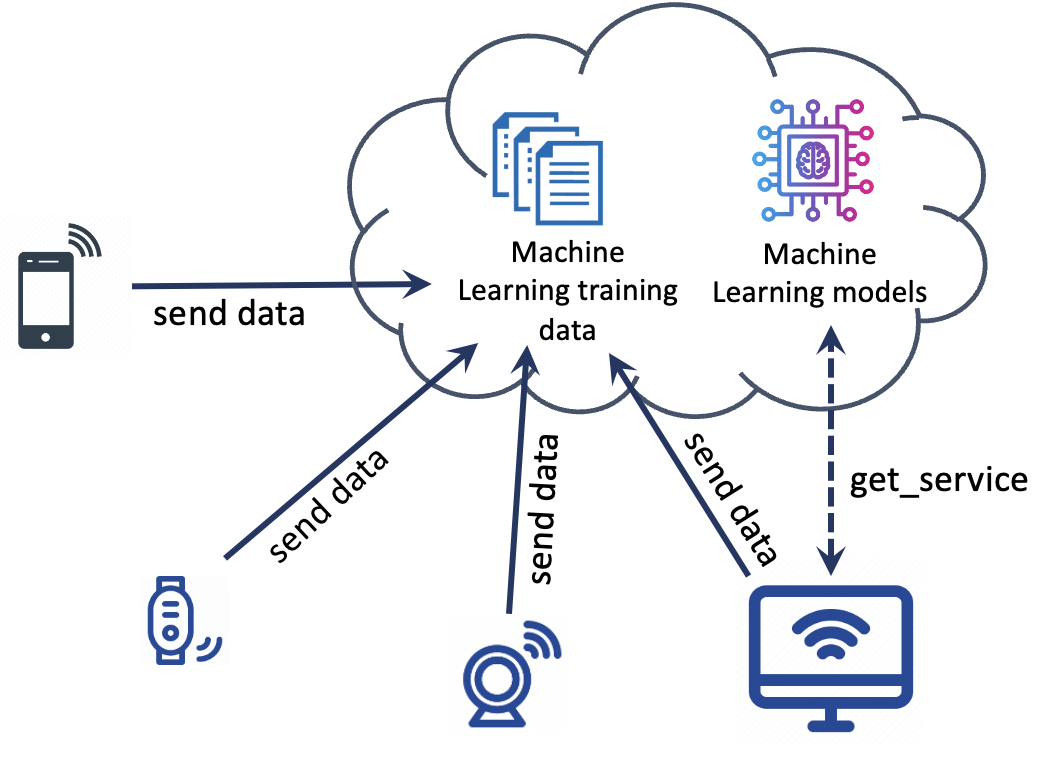}
    }
    \caption{Centralized ML Architecture\label{fig:centralizedML}}
\end{figure}

The classical architecture, illustrated in Figure \ref{fig:centralizedML}, is called Centralized Learning. In this strategy, generated data is continuously streamed into the Cloud, where high-performance servers can process them and train models efficiently. Examples of the use of such an approach are provided by popular ML-As-A-Service providers, such as Amazon Web Services\footnote{https://aws.amazon.com/}, Google Cloud\footnote{https://cloud.google.com/}, and Microsoft Azure\footnote{https://azure.microsoft.com/}.

In Centralized Learning, data is sent to the Cloud, where the ML model is built. A user uses the model through an API by requesting access to one of the available services. Within this architecture, abundant interactions generate a substantial volume of data. This can lead to privacy issues, latency, as data could be transmitted far away from the central server, and, consequently, high transfer costs.

\begin{figure}[ht]
	\centerline{
        \includegraphics[scale=0.4]{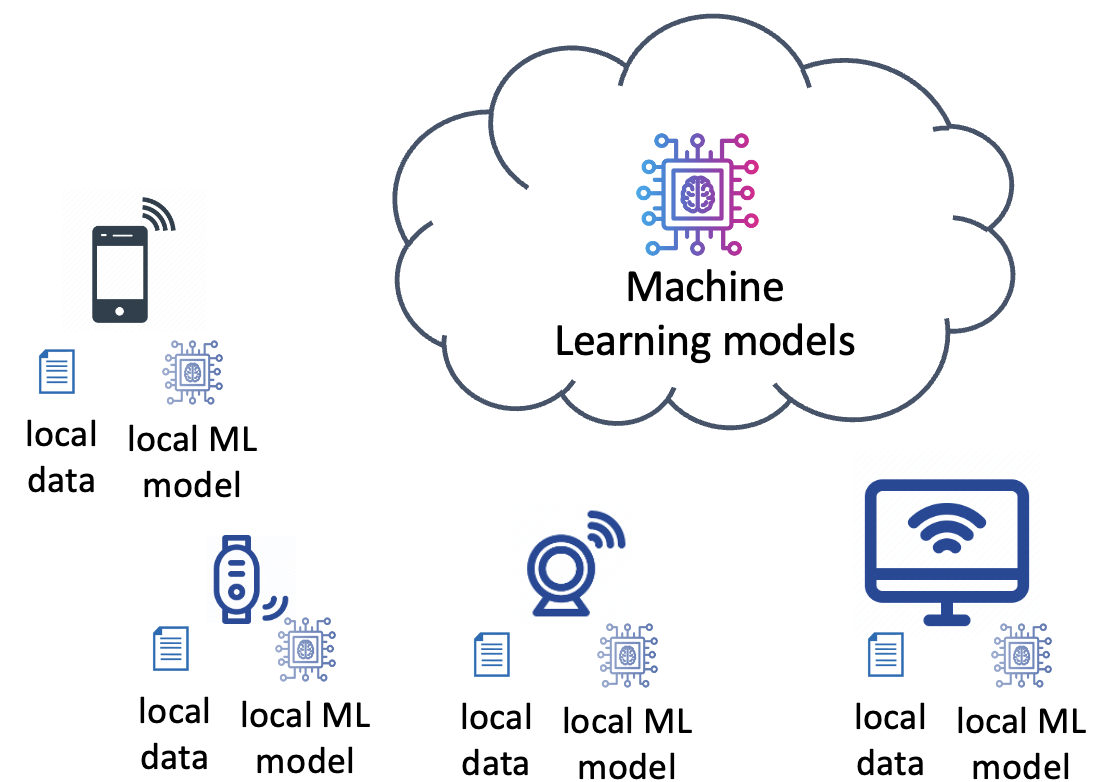}
    }
    \caption{Distributed On-Site ML Architecture\label{fig:distributedML}}
\end{figure}

Some ML tasks are moved to clients with powerful resources to overcome such drawbacks. This more recent strategy, visible in Figure \ref{fig:distributedML}, is called Distributed On-Site Machine Learning architecture. Here, each device owns a local dataset through which it can build its model. After the first interaction with the Cloud to distribute a pre-trained or generic model to the devices, no more communication with the Cloud is needed. 
Hence, privacy is obtained as data does not leave its hosts.
Although popular applications benefit from this architecture, such as medical solutions \cite{lee2013low} and smart classrooms \cite{pacheco2018smart}, models are local, and, therefore, they cannot take advantage of the results of their peers.

\begin{figure}[ht]
	\centerline{
        \includegraphics[scale=0.35]{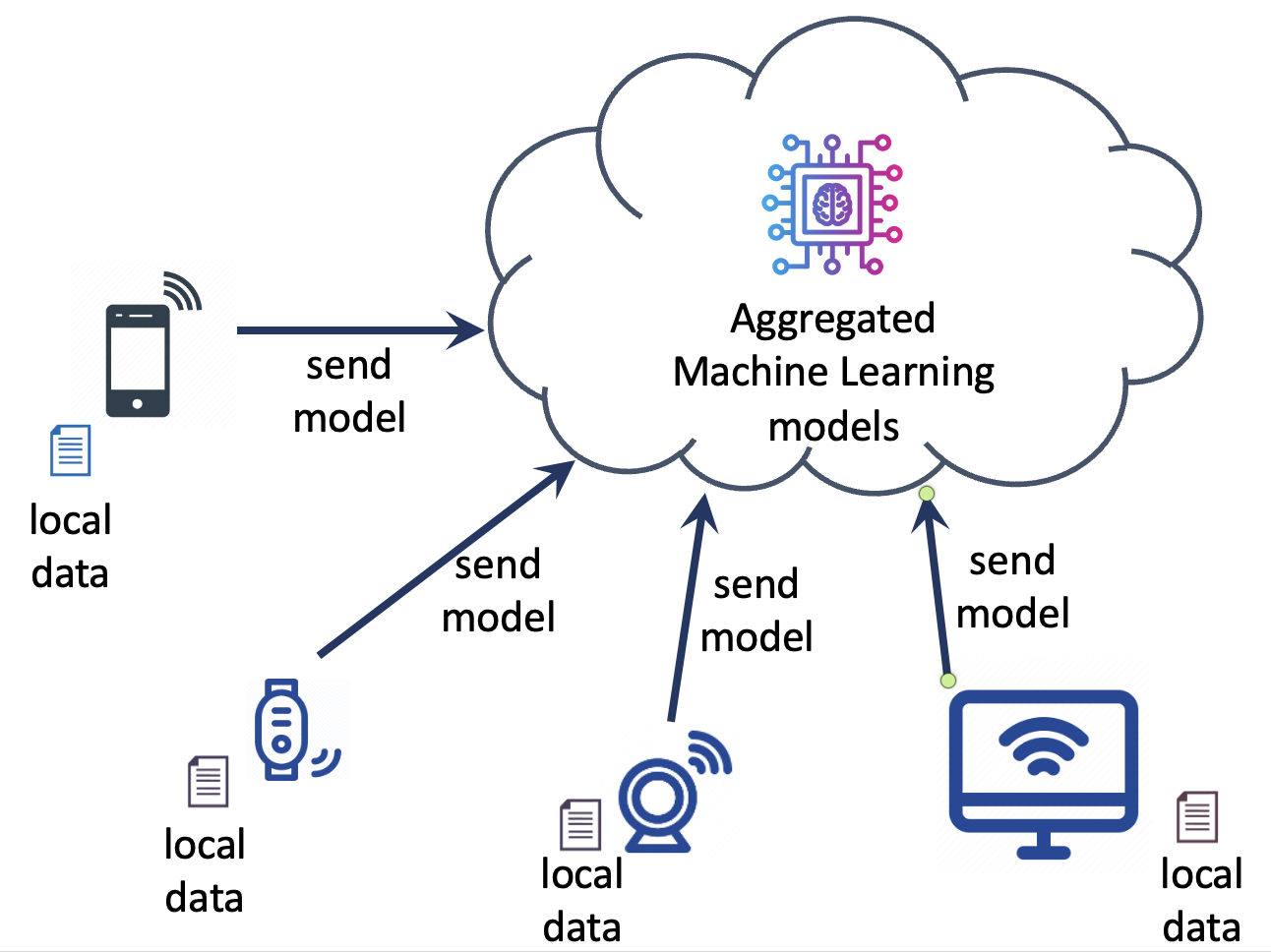}
    }
    \caption{Federated Learning Architecture\label{fig:FederatedML}}
\end{figure}

In Federated Learning, shown in Figure \ref{fig:FederatedML}, each device trains a local model leveraging local data and sends its parameters to the central curator for aggregation. Data is kept on-device, and knowledge is shared with peers through an aggregated model. In this way, FL combines all the advantages of the previous architectures. Indeed, it maintains data privacy while minimizing communication overhead by keeping raw data on devices and aggregating local model updates.

\subsection{Overview of Federated Learning}
\label{sec:overviewFL}

In the next sections, we deal with the key notions related to FL and its workflow. Finally, we discuss the primary categorizations and the main challenges inherent to such an approach.

\subsubsection{Main Concepts and Workflow}

As stated, FL is a Machine Learning strategy that allows a model to be trained across decentralized devices or servers holding local data samples while maintaining the localized data. This technique is beneficial if data cannot be efficiently centralized due to privacy regulations, network constraints, or large data volumes. 

The participants in the protocol are mainly divided into two types: devices known as ``clients'' or ``workers'' devices (e.g., IoT devices or remote servers) and a central server called ``aggregator''. Workers are individual devices, such as smartphones, IoT devices, or remote servers, that announce to the server that they are ready to run local training and participate in an FL task. Every client possesses its local dataset and utilizes it to train a dedicated local model. On the contrary, the central server or aggregator acts as the coordinating entity overseeing the Federated Learning process. The basic FL workflow consists of the following steps \cite{zhang2021survey}:

\begin{itemize}
    \item Model initialization: A global ML model is initialized on a central server or node, commonly with random parameters. During this phase, workers (e.g., IoT devices or remote servers) are selected to participate in the FL process. 
    \item Local model training and upload: Clients download the current global model to their local devices. Then, they perform local training using their data, which is kept private and not shared with the central server or other clients. The local training typically involves multiple iterations of gradient descent, back-propagation, or other optimization methods to improve the local model's performance. Following the local training, every client computes the model parameter updates and transmits them to the central server either in an aggregated form or encrypted.
    \item Global model aggregation and update: The central server collects and aggregates the model parameter updates from all the clients. The central server can employ various aggregation methods like averaging, weighted averaging, or secure multi-party computation to incorporate the received updates from each client. This process enhances the performance of the global model by integrating diverse insights from the individual client models.
\end{itemize}

\begin{figure*}[ht]
	\centerline{
        \includegraphics[scale=0.45]{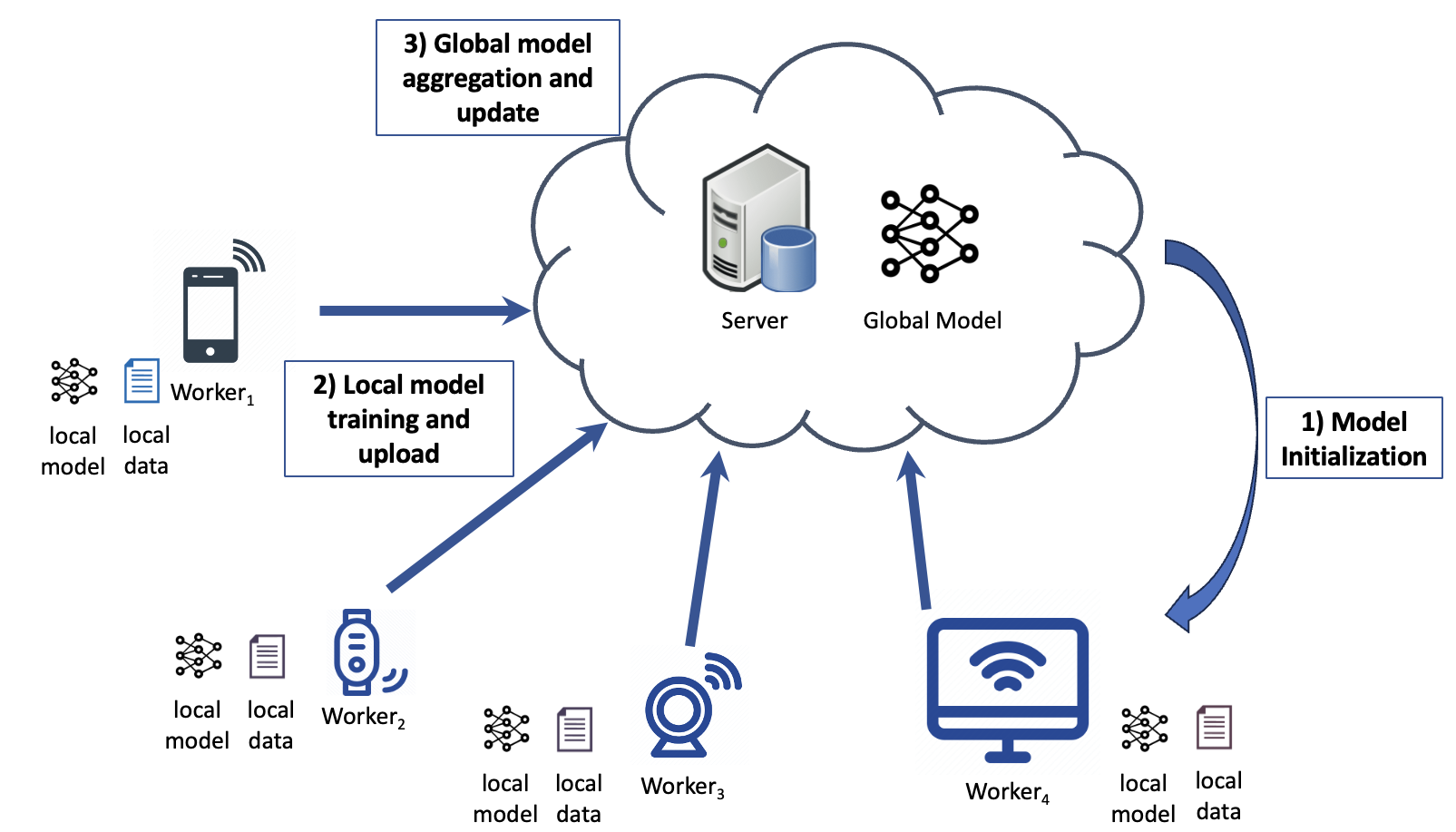}
    }
    \caption{A schematic diagram of the Federated Learning workflow \label{fig:workflowFL}}
\end{figure*}

Figure \ref{fig:workflowFL} illustrates a schematic diagram of FL workflow with the three phases described above. Observe that the last two steps of {\em (i)} iterative process of local model training and upload and {\em (ii)} global model aggregation and update are iterated across multiple epochs, continuously enhancing and refining the global model.

\subsubsection{Categorization of FL}

This section deals with the different architectures for Federated Learning based on the feature and sample spaces shared by the workers and the aggregating server.

\vspace{.2cm}

\begin{figure*}
    \centering
    \subfigure[Horizontal FL]{
        \includegraphics[width=0.3\linewidth]{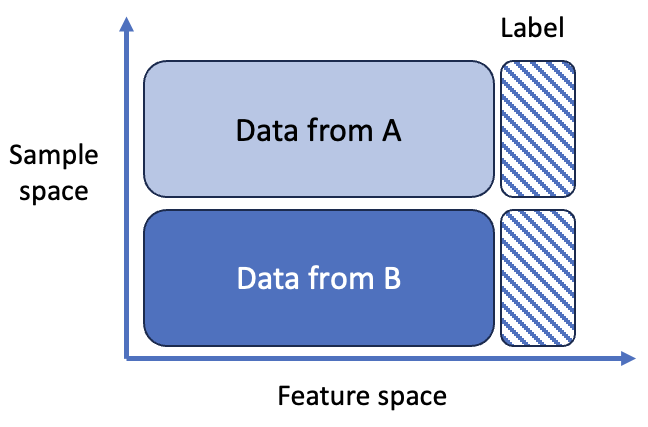}
        \label{fig:first}
    }
    \hfill
    \subfigure[Vertical FL]{
        \includegraphics[width=0.3\linewidth]{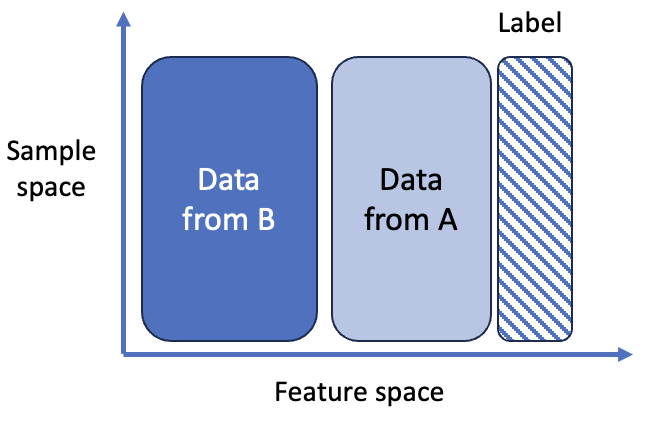}
        \label{fig:second}
    }
    \hfill
    \subfigure[Federated Transfer Learning]{
        \includegraphics[width=0.35\linewidth]{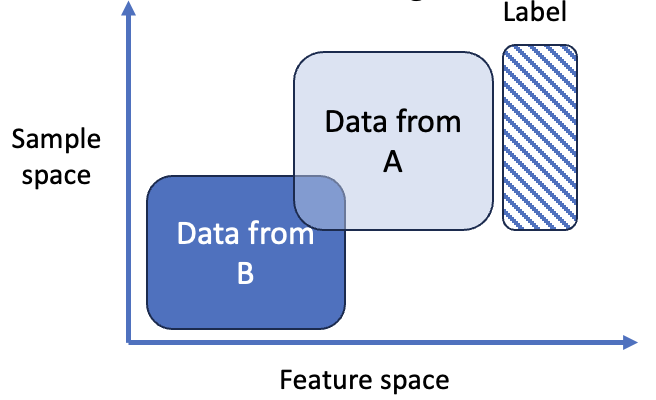}
        \label{fig:third}
    }
    
    \caption{The three categories of FL divided for feature and sample spaces}\label{fig:HorizontaleVertivalFTL}
\end{figure*}

\noindent
\textbf{Vertical FL, Horizontal FL, and Federated Transfer Learning.} A different perspective to classify FL relates to how data is distributed among the participating parties in the feature and sample spaces \cite{cheng2020Federated,zhang2021survey}.
According to this criterion, FL can be divided into Horizontal FL (HFL), Vertical FL (VFL), and Federated Transfer Learning (FTL).
Figure \ref{fig:HorizontaleVertivalFTL} shows a graphic representation of the three FL categories.

\begin{itemize}
    \item Horizontal FL refers to scenarios where the parties share the same feature space but have different data samples. This schema can be also referred to HFL as sample-partitioned FL.
    \item Differently from HFL, Vertical FL applies to the case where the actors share overlapping data samples but differ in the feature space. We also refer to VFL as feature-partitioned FL.
    \item FTL is applicable for scenarios in which there is little overlapping in data samples and features. For instance, the case in which multiple subjects with heterogeneous distributions build models in a collaboratively way.
\end{itemize}

\noindent
\textbf{Types of data heterogeneity FL.} In this part, we consider a possible FL classification according to the types of data heterogeneity. In FL, data can be Independent and Identically Distributed~(IID) or Non-Independent and Non-Identically Distributed~(Non-IID). These characteristics refer to how data is distributed among the different clients \cite{zhu2021Federated}. In the ideal scenario of an IID data distribution, the data on each client is assumed to be drawn from the same underlying probability distribution. In real-world scenarios, instead, data on different clients may have different statistical properties, feature distributions, label distributions, and data sizes. In FL, data heterogeneity refers to variations in data distributions among participants. During each federated iteration, participants are selected randomly to undertake the supervised task, incorporating features $x$ and labels $y$. Subsequently, the local data distribution of the chosen client, denoted as $P_i(x, y)$, is utilized to extract feature-label pairs from $(x, y)$. In particular, the authors of \cite{ma2022state} consider the following categories:

\begin{itemize}
    \item {\em Feature distribution skew}. It consists of an imbalance or non-uniformity in the distribution of features (input variables) across different devices, clients, or participants. Specifically, this happens when the distribution  $P_i(x)$ of the features varies from participant to participant, but the distribution of the probability $P_i(y|x)$ is the same.
    \item {\em Label distribution skew}. It means that the distribution of labels $P_i(y)$ is different for different participants, but given $P_i(x|y)$ is the same. Label distribution may vary across participants even when they share the same label annotations. For example, consider two participants, denoted as $i$ and $j$, containing data from the Fashion-MNIST dataset. In the participant $i$'s dataset, 80\% images, while the remaining 20\% display other image types. Conversely, participant $j$'s data illustrates that 85\% of the images are shirts, and the remaining 15\% depict the other types. Consequently, the distribution of labels ($(P_i(y))$) differs among participants.  However, focusing explicitly on images featuring shirts $(y = 6)$, the probability of the associated features $x$ portraying a shirt remains roughly equal for both participants. Hence, the $P_i(x | y)$ distributions are similar.
    \item {\em Quantity skew}. It is a common situation that causes data to deviate from a homogeneous distribution, and it refers to the significant difference in the quantity in different participant data~$P_i(x, y)$. For instance, participant $i$ has 500 samples, and participant $j$ has 30,0000 samples for training. Therefore, the distribution of $P_i(x, y)$ differs significantly.

\end{itemize}

\noindent
\textbf{Cross-device and cross-silo FL.} 
A further strategy to classify FL approaches is based on the participating clients and the training scale. According to this principle, FL can be divided into cross-device FL and cross-silo FL \cite{kairouz2021advances}. 

The first group consists of clients that are small
distributed entities (e.g., smartphones, wearables, and IoT
devices) holding few local data. Hence, to obtain good performance, many clients usually need to participate in the training process. Unlike the previous group, cross-silo FL clients are typically big companies or organizations (e.g., hospitals, transportation companies, and banks). In these environments, the number of participants is small (typically 2 to 100 clients), but each client usually participates in the entire training process.

\subsubsection{Primary Challenges to FL}

Most of the scientific papers focusing on FL \cite{zhao2020privacy,li2020Federated,niknam2020Federated} investigate several core open challenges that still need to be addressed, such as:

\begin{itemize}
    \item {\em Privacy protection}. One of the primary aims of FL is to guarantee the privacy and protection of data in ML solutions. It is essential that FL model training does not reveal users' private information. Most recent approaches often provide privacy at the cost of reduced model performance or system efficiency. 
    \item {\em Security}. FL systems must be robust against adversarial attacks or clients with malicious intent. FL has been analyzed through an adversarial lens to study the vulnerability of the learning process to model-poisoning adversaries \cite{bhagoji2019analyzing}. Since FL, in its classical form, is susceptible to adversarial attacks, poisoning resilience defense mechanisms should be investigated.
    \item {\em Data shortage}. ML algorithms usually demand extensive data for optimal performance, but in a distributed context (such as IoT), involved devices have limited data. Hence, FL needs local data utilization for training on each device, after which the resulting local models are sent to the server and consolidated into a global model.
    \item {\em Statistical heterogeneity}. Clients may have different data distributions, and data held by these devices may be non-IID. This makes it difficult to create a globally useful model that performs well on all clients.
    \item {\em Expensive communication}. Transmitting model updates between clients and the central server can be resource-intensive, especially in scenarios with high latency or limited bandwidth. Reducing both the total amount of communication rounds and the size of transmitted messages at each round are two main aspects to be considered.
    \item {\em Systems heterogeneity}. The presence of heterogeneous devices in terms of storage, computational, and communication capabilities leads to several challenges related to dropped devices in the network, a low amount of participation in the FL framework, and the design of scalable and flexible solutions.
    \item {\em Algorithm convergence}. The work presented in \cite{wang2019adaptive} describes a theoretical analysis of the convergence bounds of the gradient descent-based FL for convex loss functions. Anyhow, further studies on the optimum number of local workers and on the frequency of local updates and global aggregation to improve model performance and resource preservation should be deeply investigated \cite{niknam2020Federated}.
    \item {\em Lack of incentive mechanisms}. Limited research acknowledges that participants in Federated Learning lack incentives to share their data and train models. As a result, task requesters face challenges in identifying and choosing trustworthy participants with high-quality data~\cite{kang2019incentive,zhang2021incentive}. 
\end{itemize}

\subsection{Overview of Blockchain}
\label{sec:Blockchain}
\begin{figure*}[ht]
	\centerline{
        \includegraphics[scale=0.6]{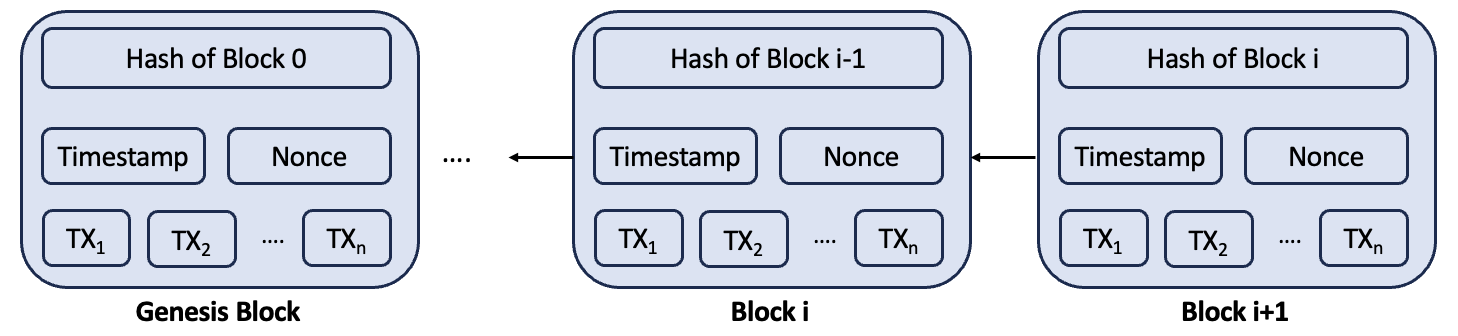}
    }
    \caption{Example of a Blockchain \label{fig:Blockchain}}
\end{figure*}

This section is devoted to providing a background description of the Blockchain technology. In the next subsections, we will describe the main concept and workflow, the strategies to build the consensus mechanism, the smart contract technology, and the main Blockchain categories.

\subsubsection{Concepts and Workflow}
In 2008, Nakamoto introduced the revolutionary Bitcoin cryptocurrency \cite{nakamoto2008bitcoin}, which operates as a decentralized and transparent peer-to-peer system. Blockchain, the underlying technology supporting Bitcoin, finds extensive utility across many financial and industrial applications due to its remarkable characteristics. A Blockchain network's most prominent feature is its utilization of a publicly digitally distributed and immutable ledger of blocks, which is shared across all participants in the peer-to-peer network without relying on any centralized trusted third party \cite{nofer2017Blockchain}. Each participant in the Blockchain network retains an individual copy of the distributed ledger to ensure data integrity, and every block contains the previous block's hash and comprises multiple transactions, as illustrated in Figure \ref{fig:Blockchain}.

\begin{figure*}[ht]
	\centerline{
        \includegraphics[scale=0.6]{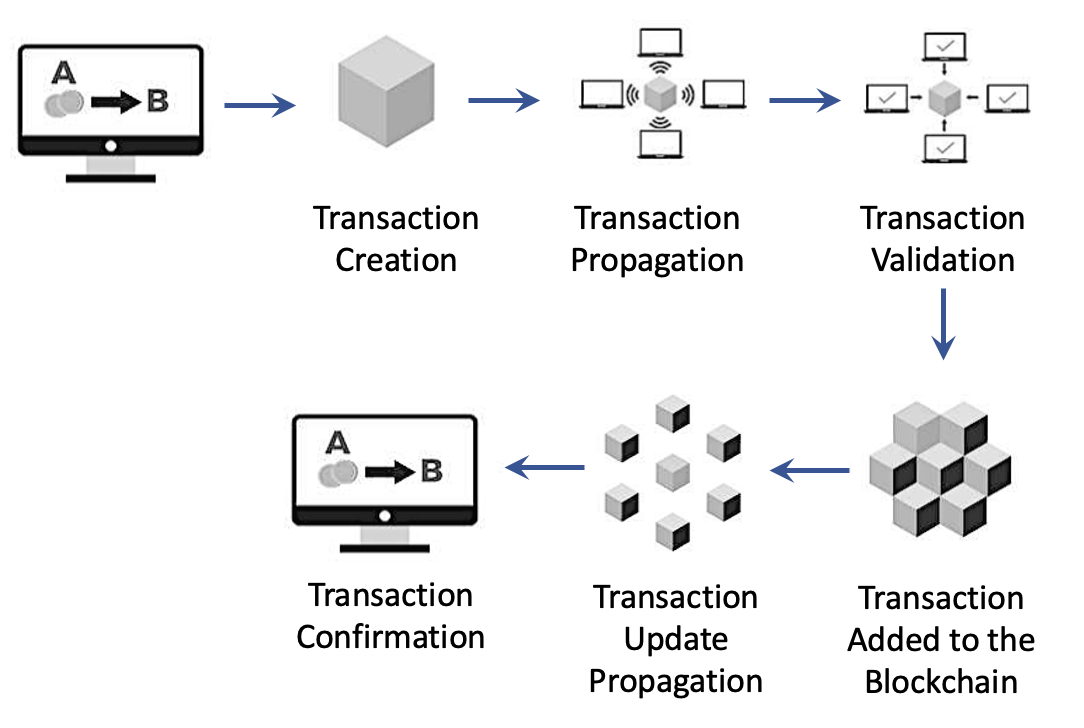}
    }
    \caption{Transactions workflow in Blockchain \label{fig:schemeTransaction}}
\end{figure*}

In addition to the transactions and the hash value of the previous block, each block includes a timestamp and a nonce, which is a random number for verifying the hash. Since hash values are unique, changes on any block in the chain would immediately change the respective hash values. Indeed, once generated, the information within each block cannot be altered, ensuring the network's immutability.
Whenever a new transaction is generated, it undergoes validation and verification through a consensus protocol carried out by {\em miners}. If the majority of nodes in the network agree by a consensus mechanism on the validity of transactions included in a  new block and on the validity of the block itself, this block is created and seamlessly integrated into the distributed ledger.
In summary, as shown in the scheme illustrated in Figure \ref{fig:schemeTransaction}, once created by a client, a transaction goes through several steps \cite{belotti2019vademecum}, namely:

\begin{itemize}
    \item {\em Propagation}. The transaction is propagated in a block towards the validating peers.
    \item {\em Validation}. The transactions collected in blocks must address the different phases of the consensus mechanism. Thereafter, the block of transactions can be attached to the Blockchain.
    \item {\em Update Propagation}. The valid transactions block is propagated throughout the network to let all nodes update their own replica.
    \item {\em Confirmation}. The consensus procedure comes to an end, and the nodes have to agree on a single chain of blocks. Blocks of transactions are published on the Blockchain and are confirmed in the final version of the ledger, from which they may no longer be discarded. 
\end{itemize}

The key features of this strategy can be summarized as follows:

\begin{itemize}
\item {\em Security}. Blockchain employs advanced cryptographic procedures to keep data secure. Once a transaction is written in a block, altering or deleting it is impractical. This makes Blockchain robust against attacks, such as fraud or tampering.
\item {\em Decentralization}. Blockchain distributes data across a network of nodes. These nodes work together to validate and record transactions. In this way, all the drawbacks of centralized solutions can be avoided, i.e., bottleneck servers or high latency due to excessive resource contention. 
\item {\em Transparency}. All transactions on a Blockchain are publicly available to all the participants in the network. This transparency can help build trust among users.
\item {\em Immutability}. Due to cryptographic hashing and chaining of blocks, once a block is added to the Blockchain, it cannot be changed, and the user cannot revoke it.
\end{itemize}

\subsubsection{Consensus mechanism}
As already stated in the previous section, Blockchain uses consensus mechanisms, such as Proof of Work (PoW) or Proof of Stake (PoS), to agree on the validity of transactions and ensure that all nodes in the network have a consistent view of the shared ledger \cite{belotti2019vademecum}.

PoW is the older and more widely adopted consensus mechanism, it has been adopted in Bitcoin and many other cryptocurrencies \cite{jakobsson1999proofs}.
To validate a block in the PoW approach, miners should find a hash value of the block that meets a certain difficulty requirement as a mathematical puzzle. The winner of this competition can validate the block of transactions and is rewarded with cryptocurrency. PoW does not guarantee consensus finality; transactions can be considered as confirmed only when included in the longest chain. PoW is designed to consume a high amount of energy because of the miners' energy-intensive computations needed to solve puzzles.

On the contrary, PoS has recently gained popularity because it is a less energy-consuming alternative to PoW, and it is used in cryptocurrencies like Ethereum 2.0\footnote{https://ethereum.org/}\cite{kovst2018transition}. PoS is adopted by a category of Blockchain algorithm where the consensus is achieved by stakes (e.g. digital assets) in the network.
Validator nodes, which are participants who hold and ``stake'' a certain amount of cryptocurrency, are chosen in a deterministic and pseudo-random manner to create new blocks and validate transactions based on the amount of cryptocurrency they hold and are willing to ``stake'' as collateral. Validators are incentivized by earning transaction fees.

\subsubsection{Smart Contract}
Smart contracts, serving as executable codes, embody a mutual agreement between two or more parties. They operate atop Blockchain to enforce and execute agreements among parties that might lack trustworthiness. These contracts define the rules, conditions, and actions to be taken when certain conditions are satisfied \cite{buterin2014next}. Moreover, it stores information, processes inputs, and writes outputs thanks to its pre-defined functions. Smart contracts are replicated on each node of the Blockchain network to prevent contract tampering.
Platforms like NXT\footnote{https://nxtdocs.jelurida.com/Nxt\_Whitepaper}, Ethereum, and Hyperledger Fabric \cite{androulaki2018hyperledger} are Blockchain-based development frameworks able to provide smart contracts to execute automatically events and actions.

Once deployed on a Blockchain, a smart contract operates autonomously. Usually, it is initiated by activating its constructor function via a transaction submitted to the Blockchain network. Each contract will be assigned to a unique address of 20 bytes. This constructor function is, then, executed, and the resulting smart contract code is permanently stored on the Blockchain. Once deployed, the creator of the smart contract receives essential parameters (e.g., the contract address). Subsequently, users can trigger any accessible functions within the smart contract by initiating transactions \cite{khan2021Blockchain}.

There are two main groups of smart contracts, namely, deterministic and non-deterministic. The smart contracts that belong to the first type do not require any information from an external party outside the Blockchain. Instead, a non-deterministic smart contract depends on information (called oracles or data feeds) from an external party.

\subsubsection{Blockchain Categorization}
Depending on the characteristics of the Blockchain, researchers and industries have defined several categories.

\noindent
\textbf{Private and Public Blockchain.}
Both public Blockchain and private Blockchain networks are decentralized and shared among their clients to register all peer-to-peer transactions without the presence of a third-party authority. However, private Blockchains are restricted to authorized participants, and a centralized entity controls access. This leads to a very high transaction processing rate with few authorized participants. Moreover, a shorter time is required to get the consensus for the network, and more transactions can be processed within a time unit.
Public Blockchain, on the other hand, is an open and permissionless network accessible to anyone, posing a risk to information privacy. However, since each transaction is open for the public to verify, they are very transparent, and the risk of hacking and data manipulation is lower when compared to private Blockchains. For this reason, it can be stated that public Blockchains are generally more secure \cite{yang2020public}. A further type of Blockchain is represented by \textbf{Hybrid Blockchain} that combines elements of both public and private ones. Some parts of the network are public, while others are private. Finally, \textbf{Consortium Blockchains}, like hybrid Blockchains, have private and public features, but they involve various organizational members working together on a decentralized network.

\noindent
\textbf{Permissioned and Permissionless Blockchain.}
Permissioned Blockchains usually involve a consortium of organizations where transactions are grouped, accessed, and verified by authorized gatekeepers instead of anonymous miners. Their implementation is arising within the finance sector \cite{grover2019diffusion}.
On the contrary, permissionless Blockchains, typically associated with public Blockchains, are open for anyone to join and participate without demanding prior authorization. They represent the first and oldest Blockchain development, in which the hashing of blocks of transactions relies on the work of many anonymous miners competing to solve a complex mathematical algorithm for that block of transactions via trial and error \cite{helliar2020permissionless}.

\section{State-of-the-art: Integration of FL and Blockchain}
\label{sec:FLandBlock}

\par Blockchain is a promising technology, providing robust and secure solutions for various applications, even when dealing with untrusted entities. In FL, it primarily safeguards user privacy. Consequently, the amalgamation of FL and Blockchain, known as Blockchain-enabled Federated Learning~(BCFL), enhances privacy and security in various distributed applications. These applications span sectors such as healthcare, cyber-physical systems, secure vehicular networks, pharmaceuticals, Industrial Internet of Things, and telemedicine\cite{lu2020Blockchain,xu2023Blockchain,otoum2020Blockchain}. BCFL effectively tackles the challenges associated with the FL paradigm by providing a range of valuable features. These include robust authentication and traceability, enhanced privacy, reliable availability, scalability, resilience against byzantine faults, resilience against inference attacks, long-term persistence, and anonymity\cite{qu2022Blockchain}.
\subsection{Benefits and characteristics of Blockchain-enabled FL}

In this section, we explore the advantages of incorporating Blockchain in the FL process. The primary limitation of current FL systems is their dependence on centralized processing, which introduces vulnerabilities such as single-point failure and susceptibility to attacks~\cite{qu2020Blockchained,lu2019Blockchain,awan2019poster}. Additionally, the extensive participation of edge devices contributes to network strain, leading to concerns about bandwidth availability and scalability~\cite{xu2023Blockchain}. Also, Blockchain technology offers a solution by providing decentralization, replacing the central server in FL applications with smart contract execution, enhancing security, and reducing the risk of malicious activities~\cite{kim2019Blockchain,li2020Blockchain}. The decentralized nature of Blockchain, preventing any single entity from having control over the entire network, aligns seamlessly with the principles of FL. In FL, data remains on individual devices and only updated models are exchanged, thereby significantly enhancing security and privacy. Nguyen \textit{et al.}~\cite{nguyen2021Federated} explored combining FL and Blockchain to create a decentralized, secure, privacy-enhanced intelligent edge network.
\par Furthermore, smart contracts automate and enforce governance rules in FL, ensuring participants adhere to predefined agreements and offering automated and transparent incentives for participants, miners, or validators based on their contributions. These agreements authenticate node contributions, perform global model computations, and facilitate node incentives based on their performance, enhancing the collaborative learning process's efficiency, audibility, and reliability~\cite{ruckel2022fairness,awan2019poster}. Incentives provided through smart contracts enhance the security and functionality of the Blockchain infrastructure while maintaining transparency and accountability~\cite{qi2021Blockchain,qammar2023Blockchain}.
\par Transactions in BCFL enable participants to trace and verify the complete history of model updates, fostering a culture of accountability within the system~\cite{lo2022toward}. Also, Blockchain's standardized protocols enhance interoperability in BCFL, allowing for seamless integration across various platforms and devices.
\subsection{The general architecture of Blockchain-enabled FL}
\begin{figure*}
    \centering
    % \floatsetup{heightadjust=object,valign=c}
    \includegraphics[scale=0.405]{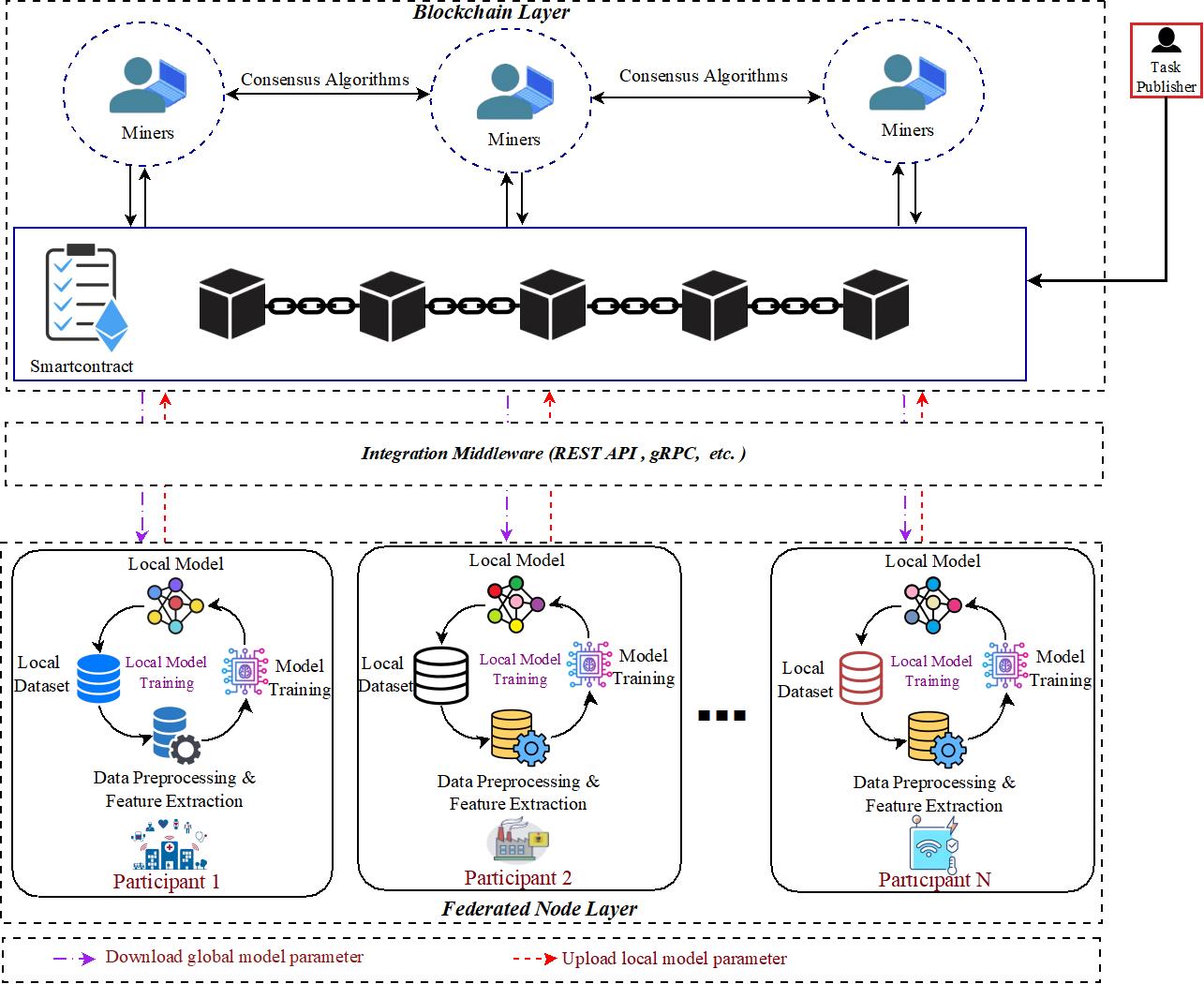}
    \caption{General architecture for the Blockchain-enabled Federated Learning.} %The process involves the task publisher submitting the initial global model to the chain. Participants' devices then download the generated block to compute the global model locally. Local updates from participating devices are uploaded to the Blockchain after local model training. Miners verify the local models, and once validated, these local models are aggregated and stored in a block during each communication round. Subsequently, miners execute the mining process on the accumulated block.}
    \label{fig:architecture}
\end{figure*}
The general abstracted architecture for Blockchain-enabled Federated Learning~(BCFL) is illustrated in Figure~\ref{fig:architecture}. This architectural framework comprises three distinct layers: the Federated node layer, integration middleware, and the Blockchain layer. In decentralized applications on a blockchain, the task publisher, as an entity or user, initiates and creates tasks, actively defining and structuring them within the blockchain. This role is pivotal in task creation and execution, starting by formally requesting a specific Federated Learning task and publishing the details into the blockchain. Subsequently, participants expressing interest in the FL tasks retrieve the models from the Blockchain and contribute their trained models back to the Blockchain. The Blockchain then operates as a central server, employing smart contracts that aggregate the models from participants. A designated miner executes this operation and creates the new global FL model to fulfill the specific FL task. In the subsequent sections, we will briefly introduce each component in detail.
\par \textit{Task Publisher}: The task publisher initiates the process by formally submitting a request for a specific Federated Learning task, meticulously outlining the parameters, requirements, and objectives. This encompasses the task publisher's identity, initialization details (such as the Machine Learning model type), targeted performance metrics for optimization, expected processing time, and other relevant information. %This encompasses a comprehensive description of the ML or Deep Neural Network~(DNN) task, including input data format, layer count, unit quantity, loss function, Learning rate, and activation function. 
Furthermore, it encompasses additional crucial parameters, such as the task's initiation time, the number of federation rounds, the total reward amount, and other relevant details. The task publisher submits details of the Federated Learning task into the Blockchain for securely and transparently storing information for participants interested in contributing to or downloading models related to the specified task. In \cite{zhao2020privacy}, the manufacturer is a task publisher to develop a smart home system. In \cite{chen2022esb}, the task publisher refers to enterprises, research institutes, or healthcare research units aiming to acquire a medical disease detection model.
\par \textit{Federated Node Layer}: For the collaborative training of an ML model, the Federated node layer encompasses a varied group of participants, including diverse devices such as smartphones, wearables, servers, and other computing entities. Participants in the FL task download the model from the Blockchain. Each participant has their private local dataset and performs data preprocessing and feature extraction on its local dataset. Preprocessing may involve cleaning the data, normalizing, handling missing values, and extracting relevant features contributing to the model's learning process. Following this, participants individually train their models using their local datasets. 
After the training process, participants in FL produce personalized model updates specific to their datasets. Subsequently, participants submit these local model updates for verification and aggregation in the subsequent phase into the Blockchain. 
\par \textit{Integration Middleware}: The integration middleware bridges FL participants and the Blockchain. Lamken \textit{et al.}~\cite{lamken2021design} employed the Representational state transfer application Programming Interface~(REST-API) to engage with Blockchain (Hyperledger Fabric) to enable a systematic allocation of network resources for recording and incentivizing gradient uploads. Additionally, the gRPC API, a remote procedure call~(RPC) protocol developed by Google, facilitates model transfer between FL participants and the Blockchain network~(Ethereum)~\cite{qammar2023securing}. 
\par \textit{Blockchain Layer:} In the Blockchain layer, pivotal elements encompass smart contracts, miners, consensus protocols, and the underlying Blockchain networks. The smart contract, another key component in Blockchain networks, operates between parties to facilitate interactions within the decentralized system. The participants utilize smart contracts (Registration Contract) to register for FL model training, ensuring transparency and immutability of conditions. After a successful registration, the revised local model is transmitted to the miners. %Miners and FL participants establish a seamless connection, maintaining uninterrupted communication. 
The miners, encompassing personal computers, cloud-based nodes, or standby servers, willingly adopt the mining software. Their primary responsibilities involve receiving local model updates~(local weights or local gradients)  transmitted by FL participants. Furthermore, miners verify and authenticate the trained local model using the consensus algorithm, which may involve Proof of Work~(PoW), Practical Byzantine Fault Tolerance~(PBFT), Proof of Stake
~(PoS), and more. Once verified, the connected miners receive updated local models from FL participants, aggregate these models, add the new updated model into the block, and subsequently upload the block onto the Blockchain network.
\begin{figure*}
    \centering
    \includegraphics[scale=0.48]{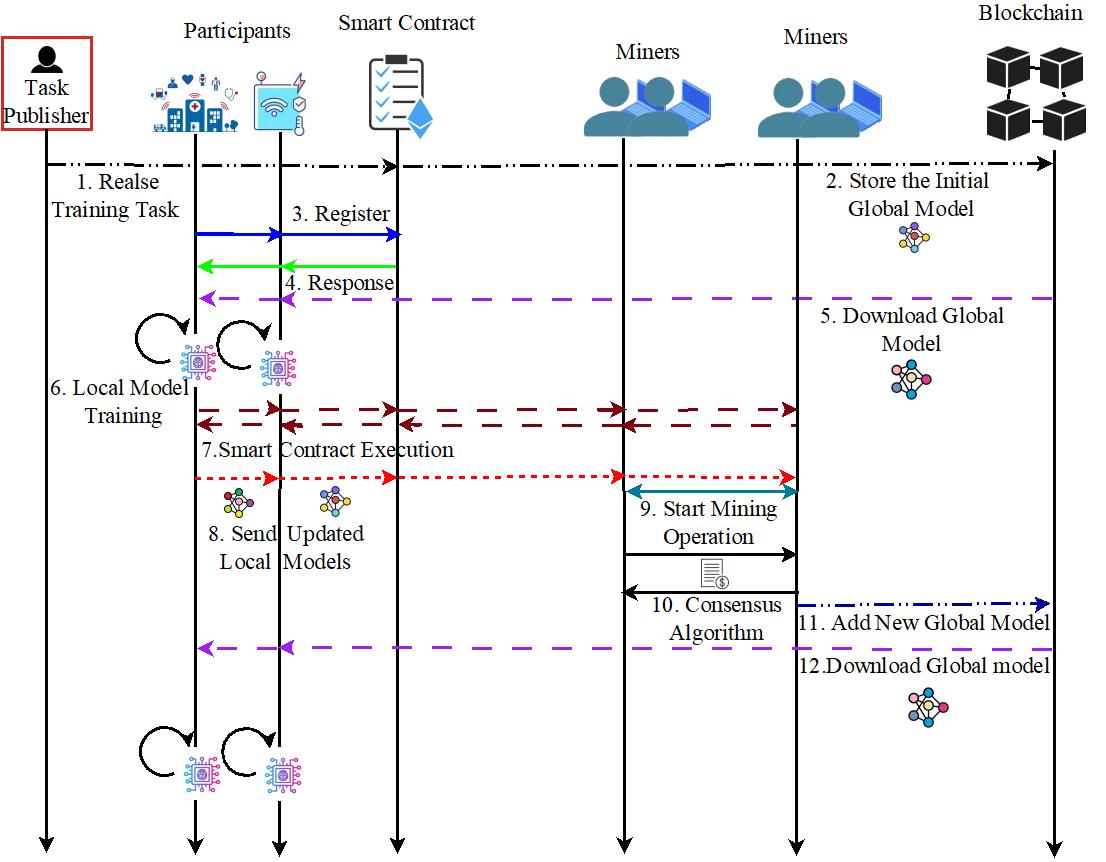}
    \caption{The high-level workflow for a single epoch Blockchain-enabled Federated Learning}
    \label{fig:workflow}
\end{figure*}
\subsection{High-level Workflow of Blockchain-enabled Federated Learning}
The high-level workflow for a single epoch in Blockchain-enabled Federated Learning is depicted in Figure~\ref{fig:workflow}. The system iterates these procedures until the model converges or attains the designated federation round.
\begin{enumerate}
    \item The task publisher initiates a service request by publishing a training task defining the parameters and details of the Federated Learning task. Following this, the task publisher deploys a smart contract to represent and regulate the Federated Learning task. This smart contract encapsulates the requisite rules, conditions, and parameters that govern the execution of the task.
    \item Then, the task publisher publishes a training task on the Blockchain.
    \item Participants willing to contribute to the Federated Learning task enroll through the smart contract. This registration process guarantees participants' adherence to the terms and conditions outlined in the smart contract.
    \item The smart contract processes the registration request, validating participant information against predefined rules. If successful, it generates a response message to acknowledge the registration, given that it fulfills the required criteria.
    \item The FL Participants download the global model from the Blockchain.
    \item The FL Participants train the model utilizing their individually preprocessed local datasets.
    \item During smart contract execution, the contract facilitates interactions among the Blockchain network, FL participants, and miners. FL participants transfer their local model updates to the miners, and the smart contract verifies the registration and validity of participants.
    \item Subsequently, participants upload their local model updates to the miners on the Blockchain. Once sufficient participants are reached, the miners, in turn, validate and authenticate these local model updates.
    \item The miners receive updates to the local models from registered Federated Learning participants and subsequently verify the received local models. 
    \item Each miner actively engages in the consensus algorithm by competitively solving complex puzzles to earn the role of a temporary leader. These temporary leaders then execute a smart contract for local model aggregation, collectively generating a new block that encapsulates information about the updated global model. Subsequently, the newly created block is disseminated to all miners within the network.
    \item Finally, a fresh block is appended to the Blockchain network, encapsulating details of the updated global model.
    \item FL participants request and download the latest global model for further training.
\end{enumerate}
The Blockchain functions as a secure and decentralized ledger, originating from the Bitcoin network, that permanently records transactions through a chain of blocks containing relevant information. 
There are two primary classifications for Blockchain storage: on-chain storage, which consolidates all records within a single ledger, and off-chain storage, where the trusted third party stores the data externally, notably through the InterPlanetary File System~(IPFS), employs a decentralized and private storage system. IPFS, a peer-to-peer distributed file system, prioritizes content-based addresses, storing hashes on the Blockchain for efficient retrieval. It offers permanent data storage, version traceability, speed enhancements, reduced bandwidth waste, and serves as a decentralized cloud storage solution, mitigating the risks associated with centralized servers. Several researchers have successfully incorporated IPFS to store actual models~(Local and Global models), ensuring immutability by sending the corresponding hash values to the Blockchain~\cite{zhao2020privacy,yuan2021chainsfl,he2021bift}. 
\section{Attacks to privacy in Blockchain-enabled FL}
\label{sec:attack}
Existing works have shown that approaches based on FL are vulnerable to attacks against data privacy. 
In particular, malicious actors can be identified both in:
\begin{itemize}
    \item the server that aims to infer sensitive information from local updates of the participants over time.
    \item the workers that can infer other participants’ sensitive information.
\end{itemize}

As a matter of fact, recent works have demonstrated that only by gradient observation a malicious attacker can successfully steal the training data, causing a deep leakage and revealing sensitive information both to a third party or the central server \cite{zhu2019deep,melis2019exploiting}.

The known attacks against data privacy that may lead to information leakage and data breaches are {\em(i)} background knowledge attack, {\em(ii)} 
collusion attack, and {\em(iii)} inference attacks \cite{qu2022Blockchain}.

\noindent
\textbf{Background knowledge attack} is a privacy-oriented attack in which an adversary leverages external information or prior knowledge to gain insights into the data used in the FL process \cite{wang2019beyond}. 
In particular, a worker regularly obtaining updates to the global model from a central authority might initiate background knowledge attacks by exploiting the differences. This results in a certain degree of privacy compromise.

\noindent
\textbf{Collusion attack} is a particular class of background knowledge attack, where malicious participants (that may be both devices or servers) collaborate to compromise the privacy of the model, leveraging background knowledge for aggregation \cite{melis2019exploiting}. If the main aim is to reveal sensitive information
to reconstruct individual data samples or learn specific patterns present in the data, this attack realizes a model inversion. Instead, if participants aim to determine whether a specific data sample is part of the training set, this attack 
is a membership inference attack. 

\noindent
\textbf{Inference attack} aims at extracting sensitive information about the data used in the training process by examining its outputs \cite{nasr2019comprehensive}. It can be divided into reconstruction attacks and tracing attacks. In the former group, adversaries try to deduce sensitive information, specific attributes, or characteristics of the training data. For instance, in \cite{li2023privacy}, a gradient inversion attack is presented. Instead, in tracing attacks, the attacker wants to determine the existence of an individual in a specific dataset.
To this last group belongs the attack proposed by Shen et al., \cite{shen2020exploiting} that exploits unintended property leakage to enable a server to infer a set of participants with target properties.
Unlike collusion attacks, where participants work together to compromise the system, inference attacks typically focus on using only information that can be deduced from the FL model's predictions or other outputs.

In the following sections, we describe recent works providing effective defenses to these attacks in scenarios involving Blockchain technology and adopting several privacy-preserving techniques while maintaining the collaborative nature of the FL paradigm.

\section{Solutions for privacy preservation in Blockchain-enabled FL}
\label{sec:solution1}

Ensuring the security and privacy of user data within the framework of Blockchain-enabled Federated Learning is a vital objective, as it necessitates a delicate equilibrium between collaborative machine learning and data protection. This section delves into various strategies and techniques to tackle the issues of safeguarding privacy in the FL empowered by Blockchain. However, these studies have collectively addressed the imperative challenge of safeguarding privacy through diverse means. Some suggested solutions involve employing homomorphic encryption,  differential privacy, secure multiparty computations, incorporating reputation-aware BCFL, and, in some instances, combining these methods to fortify privacy measures. These approaches are designed to uphold the global model's accuracy, safeguard participants' privacy, and minimize the influence of malicious local updates.  
%%%%%%%%%%%%%%%%%%%%%%%%%%%%%%%%%%%%%%%%%%%%%%%%%%%%%%%%%%%%%%%%%%%%%
\subsection{Blockchain-enabled FL Architectures for Security and Privacy Protection}

Numerous studies have introduced various approaches to incorporating Blockchain technology into Federated Learning, primarily focusing on enhancing privacy and security protection. This section explores strategies and techniques to address the challenges of ensuring privacy and security in FL empowered by Blockchain. These architectures guarantee user data protection while enabling collaborative machine learning within the Blockchain ecosystem. 
\par Furthermore, the integration offers the benefit of reducing the risk of a single point of failure attributed to the centralized aggregation curator. For example, in \cite{qu2020decentralized}, the authors introduced a BCFL system to bolster privacy and security and mitigate the risk of a single point of failure within fog computing. The study examines an attack model where adversaries attempt to manipulate training output by replacing the global model before update transmission. Achieves these goals by modifying fog servers to store global updates on the Blockchain, allowing end devices to maintain global learning models through distributed consensus, and saving only pointers on the Blockchain. In contrast, data is stored in an off-chain distributed hash table. Singh \textit{et al.}~\cite{singh2022framework} presented an alternative framework for safeguarding the privacy of IoT healthcare data through BCFL, aiming to minimize resource requirements while maintaining model accuracy and enabling fair compensation. The study also discusses the potential enhancement of this approach by not solely relying on the protocol but by incorporating a trust model and a novel consensus method within the Blockchain to support nodes. Xu \text{et al.}~\cite{xu2022privacy} presented a privacy-focused personalized reliability prediction model for IoT using Federated Learning Neural Collaborative Filtering~(FNCF), offering user privacy protection and personalized predictions, along with context awareness and improved convergence speed via local model training.
\par The authors in \cite{otoum2020Blockchain} ensure privacy and network security within vehicular networks by enabling local model training on end devices, eliminating the need to share data with the edge server. They used practical Byzantine Fault Tolerance~(pBFT) for reliable model training. The proposed framework has demonstrated remarkable performance, minimal energy consumption, low latency, high throughput, long lifetime rate, and high accuracy, approximately 97\%. Lu \textit{et al.}~\cite{lu2020low} introduced a system to reduce communication latency and enhance reliability in BCFL within edge computing. This system integrates Blockchain technology using a consensus mechanism called Delegated Proof of Stake~(DPoS) to create a decentralized training network. The system assesses latency by considering local training costs, model aggregation, parameter transmission, and block verification. It employs a deep reinforcement learning algorithm with multi-agent to optimize latency while meeting learning accuracy and bandwidth constraints. The study in~\cite{kumar2021Blockchain} presents a comprehensive framework for enhancing CT image recognition, focusing on COVID-19 detection while preserving privacy. It includes data normalization for diverse hospital data, uses Capsule Network-based segmentation and classification for precise patient identification, employs collaborative model training with BCFL, and achieves 98.68\% accuracy. The authors in \cite{guduri2023Blockchain} present a lightweight encryption strategy based on Blockchain combined with Federated Learning. This integration aims to bolster the security and privacy of electronic health records~(EHR) kept within a decentralized cloud system. The approach ensures protected access for authorized users, minimizing potential attacks on EHR data. They utilize active smart contracts to facilitate secure data transfer and validate the system's efficacy on an Ethereum-based testbed, showcasing its effectiveness. Moreover, they utilize Google Firebase to store the models.
\par 
Presently, solutions focus on favoring the selection of portable, honest local models rather than promptly and efficiently detecting Byzantine models and identifying attackers. This is mainly due to verification delays, exposing significant security risks, especially concerning untrustworthy edge and potential Byzantine attacks. To solve these issues, the authors in \cite{li2021byzantine} proposed that BytoChain enhances model verification efficiency by employing verifiers to perform parallel verification workflows and employs a consensus mechanism called Proof-of-Accuracy~(PoA) to detect byzantine attacks. It offloads the verification burden from miners by using verifiers for parallel verification workflows and introduces PoA to detect inferior models while preserving accuracy. The framework proposed in \cite{kasyap2023privacy} addresses both byzantine-robustness and inference-resistance. Utilizes permissioned Blockchain to replace the central curator, ensuring decentralized trust and fairness while protecting participant privacy. It employs private data collections in Fabric, supports multiple learning and prediction channels, and includes vertically partitioned secure aggregation to evaluate local model updates. This process calculates updated coordinate weights through Euclidean and cosine measures and determines new global model parameters via weighted averaging. Additionally, a secure prediction mechanism enables third-party applications to query the global model by securely processing raw data across peers before aggregating results for predictions.
\par Furthermore, Blockchain-based asynchronous FL aims to enhance reliability and security by introducing decentralized and transparent training processes~\cite{ali2021integration}. However, traditional Blockchain consensus algorithms are either computationally intensive or communication-intensive, hindering efficiency, and committee-based algorithms like DPoS~\cite{bamakan2020survey} may not be ideal for smart public transportation. The work of \cite{xu2022efficient} presents a novel asynchronous BCFL system tailored for intelligent public transportation, integrating a dynamic scaling factor and a unique committee-based consensus algorithm to enhance reliability while minimizing communication overhead. Specifically, the committee leader, acting as the aggregation server, identifies low-accuracy local models from its local dataset to guard against poisoning attacks. Without requiring communication and voting, a new committee leader is periodically elected from roadside units based on the latest block's hash to reduce the vulnerability to DDoS attacks.
Additionally, during the aggregation process, a dynamic scaling factor is employed to allocate suitable weights to local models based on their accuracy, subsequently improving the Learning performance of FL. Feng \textit{et al.}~\cite{feng2021bafl} introduced BAFL as a novel asynchronous strategy designed to expedite Federated Learning. BAFL incorporates two policies to enhance its workflow: one regulates the block generation rate to minimize Federated Learning delays, and the other dynamically adapts training duration to avoid transaction overloads. In contrast to the conventional FedAvg, BAFL employs an entropy weight method to evaluate device participation and records it in the Blockchain for trust. It also employs Pareto optimization to reduce model energy consumption and local device delays, striking a balance between model update speed and transaction delays. %They stored the device scores and models in a distributed ledger, which enhances FL efficiency, security, and applicability to edge devices.
 Sarhan \textit{et al.}~\cite{sarhan2022hbfl} presents a hierarchical BCFL framework for secure and privacy-preserving collaborative IoT intrusion detection. With a smart contract, transactions (model updates) and processes are executed on a secure Blockchain, enhancing system security and reliability through task compliance verification.
\par The BCFL architecture removes the necessity for a trusted server in edge environments by utilizing Blockchain to enhance trust among participants and enable all users to verify the training process and maintain transparency. The authors in~\cite{guo2022sandbox} proposed an innovative approach to facilitate collaborative learning in trustless edge computing environments. This strategy introduces a novel paradigm that includes a sandbox and a state channel for creating a secure FL environment, effectively tackling concerns regarding data privacy and quality. In addition, this approach employs smart contracts to incentivize local device and edge node participation to enhance node selection performance further. At the same time, they are utilizing a Deep Reinforcement Learning~(DRL) node selection mechanism to enhance accuracy and efficiency.  In \cite{yang2022trustworthy}, they proposed a decentralized BCFL architecture that enhances security and privacy by utilizing secure global aggregation and also employed a byzantine fault tolerance consensus protocol, which effectively safeguards against attacks from malicious servers and devices. However, the authors formulate a network optimization problem to mitigate potential long training latency that jointly considers bandwidth and power allocation. They propose transforming this problem into a Markov decision process and employing a DRL-based algorithm for adaptive and efficient resource allocation. It employs a twin delayed deep deterministic policy gradient algorithm that handles continuous optimization variables for long-term resource allocation.

%%%%%%%%%%%%%%%%%%%%%%%%%%%%%%%%%%%%%%%%%%%%%%%%%%%%%%%%%%%%%%%%%%%%
\subsection{Privacy preservation using BCFL with Differential Privacy Approach
}
%Differential Privacy is a privacy protection technique that obscures actual data with added random noise, rendering it infeasible for malicious users to infer the original data through differential attacks accurately. DP presents a well-established and effective strategy for protecting personal privacy without the high computational costs compared with the cryptographic method.by adding noise to gradient information, can challenge global model accuracy and convergence due to gradient noise.
Differential privacy incorporates randomly generated noises into data to enhance privacy and prevent precise inference of sensitive information.  In this approach, noise is introduced into the client's local parameters, ensuring the perturbation or encoding of responses independently before submission to the central curator and effectively thwarting adversaries from inferring sensitive data. However, this approach minimizes communication and computational overhead compared to cryptographic approaches. A random function $\mathcal{K}$ provides $(\epsilon, \delta)$- differentially private for $\delta \geq 0$ if, for any pair of datasets $\mathcal{D}$ and $\mathcal{D'}$ differing in at most one element, and all $\mathcal{C} \subset Range(\mathcal{K})$~\cite{arachchige2019local}.
\begin{equation}
    P[(K(\mathcal{D})\in \mathcal{C})] \leq e^{\epsilon}\times P[(\mathcal{K}(\mathcal{D'})\in C) ] +\delta
\end{equation}
The equation shows a probabilistic inequality where the likelihood of random function $\mathcal{K}$ producing a result in set $\mathcal{C}$ with dataset  $\mathcal{D}$ is limited by $e^{\epsilon}$ times the probability of obtaining a result in set $\mathcal{C}$ with a different dataset $\mathcal{D'}$, emphasizing that using $\mathcal{D'}$ increases the chance of results in $\mathcal{C}$ compared to $\mathcal{D}$ by a factor of $e^{\epsilon}$. Where $\epsilon$ denotes the privacy loss and $\delta$ denotes the error probability for the differential privacy algorithm. DP is categorized into Central Differential Privacy~(CDP) and Local Differential Privacy~(LDP). CDP relies on user trust in the data curator, incorporating random noise into the original aggregated model for privacy protection. Conversely, LDP ensures privacy without relying on trust by having individuals perturb or encode their local models. However, precise implementation of LDP is crucial to avoid inaccuracies in estimated frequencies, given that each individual independently perturbs their response~\cite{wei2020Federated}.
\par To ensure privacy in BCFL, Lu \textit{et al.}~\cite{lu2020Blockchain} suggests a novel framework for applications beyond 5G, emphasizing improved security and privacy through Blockchain integration. It also introduces a DRL optimization strategy to reduce resource costs, learning time, communication expenses, and parameter quality validation. Additionally, they added random noise in local updates for each participant to ensure privacy and tackle resource allocation challenges by optimizing resource consumption and learning quality. The system employs a DPoS protocol to validate transactions and, in the event that no consensus is achieved within a specified timeframe, diverts the computation to alternative edge servers for faster processing. Utilizes a DPoS protocol for verification, rerouting computation to alternative edge servers if consensus is not reached on time. This is followed by aggregating and verifying updates before uploading to the Blockchain, with added security through encryption using the aggregator's private key. In another work, Qi \text{et al.}~\cite{qiy2021privacy} introduced an enhanced GRU neural network tailored for traffic flow prediction. They integrate a consortium Blockchain to decentralize the FL process, ensuring that local model updates are validated by trusted consensus nodes instead of relying on a vulnerable central server. This approach effectively mitigates security risks for the central server and participating individuals. Furthermore, they implement local differential privacy by introducing Gaussian noise to local model updates, significantly bolstering location privacy protection and thwarting malicious attempts at inferring participant information through membership inference attacks.
\par
Wang \text{et al.}~\cite{wang2020Learning} proposed a secure and decentralized learning network for a mobile crowdsensing system utilizing unmanned aerial vehicles~(UAVs), allowing UAVs to securely share model updates and verify contributions without needing a central server. Furthermore, it incorporates local differential privacy to safeguard the privacy of UAVs' updated local models and maintain privacy. Additionally, it incorporates a two-tier reinforcement learning-based incentive system to encourage the sharing of high-quality models among UAVs, even when network parameters are not explicitly disclosed.
Xu \text{et al.}~\cite{xu2023Blockchain} proposed a novel BCFL model for the Industrial Internet of Things, incorporating adaptive differential privacy to safeguard local model privacy without compromising accuracy. They used the Laplace mechanism, which relies on local DP, to introduce noise into the intermediate parameters during the model update phase. The cropping threshold can adapt automatically based on the training progress, effectively minimizing the influence of additional noise on model accuracy. The model further implements model parameter validation and proof of contribution consensus to effectively detect and prevent malicious node poisoning attacks, ensuring fairness through reputation and incentive mechanisms. A node reputation system is designed to assess participant reliability, calculated using a multi-subjective logic model. It serves as the basis for consensus committee election and incentives, enhancing overall fairness among participating nodes. 
\par The authors in \cite{zhao2020privacy} designed a privacy-preserving BCFL for home appliances. They employed DP on locally trained customer models specifically to the gradient of the local model using the regularization method, and selected customers acted as miners to aggregate the model. They suggested using the IPFS for off-chain model storage to address limited storage, recording their hashes in the Blockchain. It also introduced a novel normalization technique for improved accuracy and proposed an incentive mechanism for rewarding honest customers. Attained a minimum accuracy of 90\% but highlighted the existence of a trade-off between accuracy and the level of induced noise. In \cite{wan2022privacy}, it introduces a novel approach to tackle crucial challenges within edge computing environments by BCFL with DP facilitated by Wasserstein Generative Adversarial Networks~(WGAN) in B5G networks. Minimize communication overhead between edge devices and the cloud, address data falsification concerns, and promote a collaborative data-sharing approach. WGAN generates controllable random noise that complies with DP requirements and is injected into model parameters, bolstering the privacy and security of local model data. Applying game theory to attain Nash Equilibrium among the generator, discriminator, and DP-identifier enhances the overall efficacy. In \cite{qu2022fedtwin}, they introduced a novel BCFL approach, integrating generative adversarial networks and differential privacy~(GAN-DP) for privacy and decentralization in Delay-Tolerant~(DT) networks. They used a modified Isolation Forest to detect and remove falsified local models. They employed an improved Markov decision process to select optimal DTs for flexible asynchronous aggregation. GAN-DP addressed privacy concerns and encouraged end devices to contribute sensitive data, enhancing system performance. It also supported local data augmentation, mitigating size and class balance issues, improving learning efficiency, and reducing operational costs.  Cui \text{et al.}~\cite{cui2021security} designed an innovative GAN-driven differentially private algorithm to protect the privacy of local model parameters by adding controlled noise, ensuring compliance with differential privacy requirements while improving the utility of the anomaly detection model in IoT Infrastructures. 

\par Safeguarding medical records data represents a crucial challenge in the modern digital age, demanding advanced protective measures as cyber threats evolve. In \cite{zhang2021Blockchain} and \cite{ javed2023sharechain}, proposed BCFL to enhance patient data privacy in healthcare applications along with DP noise added into the local models. The system addressed storage efficiency by storing only the hash value on IPFS within the Blockchain while the original data was kept locally.
Liu \text{et al.}~\cite{liu2023pd2s} introduced a cross-layer architecture, employing differential data sharing for origin data and model providers. Their targeted incentive mechanism, designed as a two-stage Stackelberg game, optimizes utility, enhancing privacy and speeding up performance, surpassing the simple shared model and data schemes by 1.72 and 2.59 seconds, respectively. Furthermore, Laplace differential privacy protects intermediate privacy parameters during aggregation.
Li \text{et al.}~\cite{li2021Blockchain} proposed an architecture to enhance FL privacy and security while dealing with lazy clients and SPoF issues. It introduces a bounded loss function to analyze the relationship between block creation and the impact of lazy clients on training efficiency. Optimizing the loss function improves performance despite the presence of lazy clients. Also, it provides learning incentives by optimizing computational resource allocation and ensuring data privacy through differential privacy. In \cite{ji2023lafed}, the authors have presented a lightweight authentication framework tailored for BCFL. This framework incorporates a flexible Blockchain consensus algorithm and zero-knowledge proof to validate the identity of participants. Furthermore, an adaptive model aggregation algorithm, considering both the model's quality and the contribution of each node, is employed to boost overall performance, thereby attaining a high level of training accuracy. The Laplacian mechanism for differential privacy protection is applied in intermediate gradients to protect local data privacy from inference assaults while reducing the possibility of data leaking.

\par Numerous researchers are developing custom Blockchains for various applications, including exchanging and verifying local model parameters in IoT-based Federated Learning. For instance, Salim \text{et al.}~\cite{salim2021Blockchain} developed a Python-based custom Blockchain for Blockchain-based Explainable Federated Learning~(DP-BFL) to enhance security in IoT-based Social Media 3.0 networks. DP-BFL employs differential privacy to safeguard the exchanged local model updates and the aggregated global model from potential inference or membership attacks. Furthermore, this allows Internet-enabled devices to actively contribute to a globally preserved privacy model by uploading local updates to Blockchain miners. These miners evaluate and reward these contributions, with the added feature of introducing adaptable Gaussian noise to enhance privacy. Miao \text{et al.}~\cite{miao2022intelligent} developed a secure data-sharing model using peer-to-peer FL with Blockchain-distributed ledgers to ensure data transparency and differential privacy for enhanced data privacy in IoT. They employed team-based data sharing with reward and punishment mechanisms to guarantee high-quality and reliable data sharing, where team sponsors initiate tasks and assess members' contributions, rewarding active participants and excluding poorly engaged members. They suggested a proof of model contribution consensus algorithm that relies on the contribution of the training model to enhance computational efficiency. Experimental results confirmed the effectiveness of their approach, highlighting high accuracy and improved privacy in IoT. Zhang \textit{et al.}~\cite{zhang2023privacy} present a privacy-protecting FL framework for IoT that employs Blockchain and committee consensus.
Local updates are verified through Blockchain, ensuring data privacy with local differential privacy where Laplace noise is used. Committee nodes validate model parameters, and when sufficient validation responses are received, updates are aggregated through a smart contract for the next training round. 
\par In \cite{lu2019Blockchain} explores a permissioned Blockchain system with the Proof of Training Quality~(PoQ) consensus process, optimizing node computing resources during data model training. The Laplace mechanism enhances local data model privacy and improves computing resource utilization and efficiency of the data-sharing scheme.
Chen \textit{et al.}~\cite{chen2023privacy} introduced an efficient Privacy-Preserving and Traceable FL framework with minimal overhead and high performance. Their innovative approach incorporates hierarchical aggregate Federated Learning, involving sub-aggregators and aggregators and adding noise to local model parameters using random seeds. The sub-aggregator can reconstruct pseudorandom weights with user IDs or decrypt subtracted parameters. After aggregating and encrypting the parameters, the sub-aggregator forwards them to the aggregator, which decrypts and combines parameters, subtracts user-added noise, and obtains global parameters sent to the server. In \cite{arachchige2020trustworthy}, PriModChain, a specialized FL architecture for Industrial Internet of Things networks, incorporates a differential privacy approach to add artificial noise to locally generated models, which reduces the risk of the identification of individual records. The secure transfer of the global ML model is facilitated through smart contracts, ensuring consensus on update verification and transparency in FL updates. Simulations in Python evaluate PriModChain's feasibility in terms of security, privacy, safety, reliability, and resilience, highlighting its innovative features in promoting unbiased and error-free data manipulations for enhanced FL safety and reliability against external data threats. The frameworks~\cite{chen2023privacy, arachchige2020trustworthy} integrate FL with Blockchain and IPFS, guaranteeing the traceability and immutability of model parameters, particularly suitable for Industrial Internet of Things scenarios. Table \ref{tab:dp} comprehensively outlines the strategies employed for privacy preservation in BCFL by applying the differential privacy approach. The table details the diverse methods and techniques this privacy framework utilizes to ensure robust privacy measures in FL on the Blockchain.

\begin{table*}
\centering
\caption{Privacy Preservation in BCFL using using differential privacy approaches}\label{tab:Differential_Privacy}

\begin{tabular}{|l|p{1cm}p{1.9cm}p{1.9cm}p{1.9cm}p{1.9cm}l|}
\hline
\textbf{Reference paper} & \textbf{CDP/ LDP} & \textbf{Exponential Distribution}&\textbf{Gaussian Distribution}&\textbf{Laplace Distribution}&\textbf{Random Distribution}&\textbf{Parameter} \\
\hline
\cite{lu2020Blockchain}&LDP& \checkmark&&&&Local Gradient\\
\hline
\cite{qiy2021privacy,qu2022fedtwin,li2021Blockchain}&LDP& &\checkmark&&&Local Weight\\
\hline
\cite{wang2020Learning,javed2023sharechain,liu2023pd2s}&LDP& &&\checkmark&&Local model\\
\hline
\cite{zhao2020privacy,xu2023Blockchain,zhang2023privacy}&LDP& &&\checkmark&&Local Gradient\\
\hline
\cite{wan2022privacy}&LDP& &&\checkmark&&Local Weight\\
\hline
\cite{lu2019Blockchain,cui2021security,ji2023lafed}&LDP& &&\checkmark&&Local Weight\\
\hline
\cite{salim2021Blockchain,arachchige2020trustworthy}&LDP& &\checkmark&&&Local Gradient\\
\hline
\cite{miao2022intelligent}&CDP& &&\checkmark&&Global Gradient\\
\hline
\cite{chen2023privacy}&LDP& &&&\checkmark&Global Gradient\\
\hline
\end{tabular}
\label{tab:dp}
\\
\vspace{0.3cm}
\footnotesize CDP: Central Differential Privacy, LDP: Local Differential Privacy,\\ %HbC: Honest-but-Curious,MalC: Malicious Client

\end{table*}

%%%%%%%%%%%%%%%%%%%%%%%%%%%%%%%%%%%%%%%%%%%%%%%%%%%%%%%%%%%%%%%%%%%%
\subsection{Privacy Preservation in BCFL using Homomorphic Encryption-based Approaches}

Homomorphic Encryption~(HE) is a technique that enables computations on encrypted data, yielding encrypted results without requiring data decryption~\cite{antwi2021privacy}. In FL, users can employ HE to secure their parameters while sharing them with the server, which protects data privacy and facilitates accurate model aggregation~\cite{zhang2020batchcrypt,zhang2022homomorphic}. Typically, in FL, the server involves the processing function $f$, which aggregates parameters from local models across all participating nodes. The encryption computation utilizing HE is detailed in equation \ref{eq:HE} as follows:
\begin{equation}
    E(m_1) * E(m_2)*\dots * E(m_n)=E(m_1*m_2*\dots*m_n)\label{eq:HE}
    %,\\\forall m_1,m_2,\dots,m_n
\end{equation}

Where, $(m_1,~ m_2,~ m_3, ~\dots,~ m_n)$ denotes the parameters and $E$ represents the encryption algorithm.
\par 
Chen \textit{et al.}~\cite{chen2021ds2pm} developed a data-sharing private model that utilizes BCFL. The study addresses data privacy by proposing a scheme based on FL and employs HE to safeguard user parameters during parameter updates. To alleviate storage issues and manage diverse data formats, the work combines Blockchain storage with off-blockchain key-value storage, using Blockchain only for data pointers. An innovative on-chain data retrieval mechanism selects data providers for FL. Additionally, the research introduces a consensus mechanism called contribution authorizing Byzantine fault-tolerant algorithm~(Con-dBFT), based on contribution, to improve fairness and efficiency in the system. Wang \textit{et al.}~\cite{wang2022Blockchain} proposed a BCFL to address the security threats faced by the privacy-preserving FL, which enhances Multi-Krum technology by integrating it with HE, resulting in ciphertext-level model aggregation and filtering. This method ensures the verifiability of local models and preserves user privacy. In \cite{weng2019deepchain}, it also protects the local model's gradients through encryption using the Threshold Paillier encryption algorithm.
Furthermore, it introduces a reputation-based incentive mechanism within the Internet of Vehicles to incentivize honest participation in FL, and the authors used a semi-decentralized consortium Blockchain structure with an Elliptic Curve signature and Merkle tree to ensure data security. Sun \textit{et al.}~\cite{sun2022Blockchain} proposed BCFL, which encrypts the local gradients using the Bresson-Catalano-Pointcheva~(BCP) mechanism and then adds homomorphic noise to each encrypted gradient. The modified gradients are then gathered and assessed for quality using a joint audit algorithm. The system identifies any gradients that lead to the global model's degradation, effectively removing them from the model. It then aggregates the remaining gradients, generating a new global model with reduced processing time. However, the behavior and audit chains may become overwhelming as data owners increase, leading to delays and processing times, potentially limiting its practical use in large-scale Federated Learning scenarios. In another work, Miao \textit{et al.}~\cite{miao2022privacy} created a BCFL-based byzantine robust model to ensure privacy and mitigate the system to infer the client's local data. They create a reliable global model by identifying malicious gradients and honest gradient vectors through cosine similarity.
Additionally, they used the Cheon-Kim-Kim-Song~(CKKS) scheme based on fully homomorphic encryption to safeguard privacy and encrypt local gradients. Furthermore, it significantly decreased the computation and communication overheads. In \cite{qammar2023Blockchain}, researchers utilized a similar approach to safeguard the local model from inference attacks. Chen \textit{et al.}~\cite{chen2022esb} also an effective non-interactive designated decryptor function encryption method as a novel lightweight cryptography tool. The method effectively maintains the accuracy of the global model with comparatively low and efficient transmission costs. Sezer et al.~\cite{sezer2023ppfchain} introduced the BCFL framework to guarantee the security and privacy of IoT sensor-based structures utilizing sampled data from electrochemical sensors. Within this architecture, they employed Federated models and cryptographic primitives to ensure user and data privacy in off-chain fog nodes with high accuracy, efficiency, and security.
\par However, existing HE-based systems face significant challenges, such as the reliance on trusted third parties for key management, increased complexity and vulnerability, and scalability issues with Deep Learning~(DL) models due to computational constraints in encrypting and decrypting the trainable parameters~\cite{zhu2021distributed}. The authors in \cite{muazu2023iomt} introduced a BCFL system empowered by edge computing for resource management in the Internet of Medical Things~(IoMT). It employs an improved linear regressor model and Paillier encryption for gradient parameter security. Mobile devices act as initiators for model bootstrapping and local task initialization, while validators, selected based on computing capabilities, engage in Blockchain consensus processes, block verification, and validation. The computing threshold for validator miners is determined using maximum likelihood estimation, ensuring a data-driven approach to resource allocation. The resulting blocks are digitally signed, hashed, and encapsulated into the Blockchain, enhancing security features for IoMT and edge computing. 
\par The approach presented by Qi \textit{et al.}~\cite{qi2021Blockchain} guarantees gradient privacy using HE while tackling trust issues and Single Point of Failure~(SPoF) through a reputation system based on smart contracts. Additionally, the model addresses Blockchain storage challenges by implementing an on/off-chain storage strategy. Li \textit{et al.}~\cite{li2023privacy} proposed a privacy-preserving FL system, employing distributed ElGamal encryption to safeguard gradient inversion attacks. The system recovers original data from local sign-based quantized gradients and utilizes smart contracts for secure self-aggregation among participants without reliance on a centralized server. Some works have focused on privacy in vertical FL, proposing a novel technique that utilizes DL and Blockchain to preserve the privacy of electronic health records by developing a secure logistic regression architecture~\cite{alzubi2022cloud}.
\par In \cite{arazzi2023fully}
uses a combination of FL, Blockchain, and HE to compute a global behavioral fingerprinting model for a target object in an IoT context. This fingerprint is derived from the interactions of an object with different peers and allows anomaly detection in the network to be performed. The underlying model, thanks to HE, guarantees the privacy of both the target object and the different workers, as well as the robustness of the strategy in the presence of attacks. 
\par Li \textit{et al.} enforced privacy safeguards in \cite{li2021privacy} by combining BCFL and HE within a traceable identity-based scheme, ensuring the records' integrity and traceability. They aimed to establish an anonymous identity-based scheme for safeguarding driver identity privacy by adopting FL and utilizing the classic Feige-Fiat-Shamir zero-knowledge-proof authentication. 
\par Table \ref{tab:HE} offers a comprehensive summary of privacy preservation within BCFL, utilizing homomorphic encryption with diverse approaches. Awan et al.~\cite{awan2019poster} enhanced the Paillier cryptosystem, incorporating features like additive Homomorphic Encryption and proxy re-encryption to safeguard individual local model updates in FL. Their approach addresses issues such as random client dropouts through asynchronous recording on the Blockchain. Integrating BCFL mitigates multiparty dropout and enhances transparency, verifiability, and data privacy protection.

\begin{table*}
\centering
\caption{Privacy Preservation in BCFL using HE}
\renewcommand{\arraystretch}{1.4}

\begin{tabular}{|p{1.3cm}|p{1.5cm}p{3.5cm}p{2cm}p{2.5cm}l|}
\hline
\multirow{2}{*}\textbf{Reference paper} & \textbf{Encryption Type} & \textbf{Privacy scheme} & \textbf{Parameter}& \textbf{ Attack against}&\textbf{Adversary} \\
\hline
\cite{chen2021ds2pm}&PHE& Additive &Local Gradient&I & Server
\\
\hline
\cite{wang2022Blockchain}&PHE&  Paillier additive&Local Weights &I\&P & HbCS~\&~MalC
\\
\hline

\cite{sun2022Blockchain}&FHE&  BCP &Local Gradient &I\&P & HbCS
\\
\hline
\cite{qammar2023Blockchain,miao2022privacy}&FHE& CKKS &Local Gradient &I & Server~\&~MalC
\\
\hline
\cite{chen2022esb}&FE& NDD-FE & Local Weights&I & -
\\
\hline
\cite{muazu2023iomt}&PFE& Paillier additive & Local Gradient, Global Gradient&Transaction Hacking, I, Impersonation\& 51\% attack & Insider or Outsider
\\
\hline
\cite{li2023privacy}
&PHE& Distributed ElGamal & Local Gradient&Gradient Inversion  & HbC clients 
\\
\hline
\cite{zhou2022Blockchain}&Encryption &Proxy re-encryption, ECC, SS,CH & Local Weights&I & HbC clients 
\\
\hline
\cite{alzubi2022cloud}&Encryption & Proxy re-encryption& Global Weights&I & -
\\
\hline
\cite{awan2019poster}&PFE& Paillier additive\&Proxy re-encryption&Local Gradient &I&SHbCS\\
\hline
% \cite{qu2021proof}&FHE & -& Local Weights&Label Prediction& -
% \\
% \hline
% \cite{feng2021Blockchain}&FHE & -& Local Model&I,GAN attacks& Clients
% \\
% \hline
\cite{qi2021Blockchain,weng2019deepchain}&PHE& Paillier additive &Local Gradient&I\&P &Insider/Outsider
\\
\hline
\cite{li2021privacy}&FHE& Dijk-Gentry-Halevi-Vaikutanathan &Local Model\& Global Model &I\&P & MalC,HbCS \& SHbCS
\\
\hline

% \cite{chen2022esb}&FE& NDD-FE & Local Weights&I & -
\end{tabular}
\\
\vspace{0.3cm}
\footnotesize FHE: Fully Homomorphic Encryption, PHE: Partially Homormophic Encryption, HbCS: Honest-but-Curious Server,MalC: Malicious Client,\\ I: Inference attack, P: Poisoning attack, SS:Secret Sharing, ECC:Elliptic Curve Cryptography, CH: Chameleon hash, SHbCS: Semi  Honest-but-Curious Server
\label{tab:HE}
\end{table*}

%%%%%%%%%%%%%%%%%%%%%%%%%%%%%%%%%%%%%%%%%%%%%%%%%%%%%%%%%%%%%%%%%%%%
\subsection{Privacy preservation using  BCFL with Secure Multiparty Computation approach}
Secure Multiparty Computation~(SMPC), introduced by Andrew Yao in 1982, forms the foundational protocol for secure computations~\cite{zhao2019secure}. It facilitates different parties~$( P_1, P_2 \dots P_n)$, with private data~$(d_1, d_2 \dots d_n)$, in jointly computing an objective function~$(f)$ on their private data~$f( P_1, P_2 \dots P_n)$, thus preserving the confidentiality of the input data~\cite{antwi2021privacy}. The authors in \cite{li2020Blockchain} present BCFL with novel committee consensus, utilizing Blockchain for global model storage and local updates. The innovative committee consensus minimizes computation and enhances security. A committee validates updates in each round, reinforcing the global model while rejecting incorrect ones. It allows flexible participation, enabling nodes to join or leave without disruption, and uses Smart Contracts driven by Blockchain transactions to execute the central server functions.  
\par However, some studies emphasize persistent security concerns in key management, particularly regarding secret key ownership in adopted cryptographic systems. To tackle this issue, multiple studies, exemplified by \cite{jiang2021pflm} and \cite{fang2022privacy}, advocate for the adoption of the SecAgg protocol~\cite{bonawitz2017practical}. Within this protocol, secret keys are collaboratively shared and securely stored using Blockchain. Fang et al.~\cite{fang2022privacy} also address these concerns by employing Blockchain to verify global model gradients, effectively mitigating the potential risk of tampering attacks. Moreover, gradient compression methods are employed to alleviate communication overhead. In \cite{jiang2021pflm}, a variant of ElGamal encryption was employed to validate the accuracy of aggregated results.
\par In the architecture proposed by \cite{kalapaaking2023Blockchain}, multiple smart hospitals in different regions are assumed, each equipped with a cluster of IoT medical devices and an edge server executing FL tasks. This verification involves encrypted inference through a SMPC protocol. Upon verification, the Blockchain node obtains the authenticated portion of the local model. Utilizing SMPC-based secure aggregation, the Blockchain and the hospital collaborate to reach a consensus on the global model, which is securely stored in the Blockchain. The tamper-proof storage system then disseminates the revised global model to all involved hospitals in the Federated Learning round.
\par In a Blockchain-based decentralized, secure multiparty Learning system outlined in \cite{wang2020ai}, every client calculates and disseminates its local model via the Blockchain. Following a calibration process specifically designed for edge computing-based IoT applications, clients execute models received from other participants. The system employs a cooperative mining strategy, incorporating on-chain and off-chain mining, to address potential attacks during model broadcasting and calibration. 
%%%%%%%%%%%%%%%%%%%%%%%%%%%%%%%%%%%%%%%%%%%%%%%%%.%%%%%%%%%%%%%%%%%%
\subsection{Privacy preservation using BCFL with reward-driven approaches}
Integrating BCFL with incentive mechanisms not only addresses the challenge of preserving user privacy and encouraging active participation but also ensures the confidentiality and security of the BCFL system. By leveraging smart contracts, BCFL establishes a transparent and tamper-proof framework for fair and verifiable incentives, mitigating concerns about opaque reward structures in traditional BCFL platforms. This innovative integration promotes collaboration and significantly enhances the effectiveness and trustworthiness of the BCFL system~\cite{zhan2021survey,li2021privacy}. BCFL's selection process is guided by a strong emphasis on client reputation. Higher-reputation clients are more likely to contribute reliable and high-quality training. After each training task, client reputations are updated based on their behavior, influencing client selection in subsequent training by considering their reputation records.

\par Assessing the contributions of diverse data providers is fundamental for fair profit allocation. Implementing reasonable contribution evaluation criteria enhances the incentive mechanism, attracting more participants to join. Clients' contributions can be distilled into two main categories: data quality and data quantity. For example, Salim \textit{et al.}~\cite{salim2021Blockchain} introduced an incentive mechanism designed to combat free-riding attacks by proportionally rewarding participants based on the quality of their contributions. They implemented the Quality-Based Consensus~(QBC) algorithm in DP-based BCFL, ensuring that only legitimate local updates contribute to the global model. QBC rewards participants for added updates, promoting high-quality contributions, and selects the consensus leader based on the miner with the highest accuracy for inclusion of the most qualified models in the global update. 
\par Furthermore, Qi \textit{et al.}~\cite{qi2022high} proposed a mechanism to motivate data owners to provide high-quality data by establishing a distinct equilibrium by analyzing noncooperative games. A reputation layer utilizing Blockchain for collaborative assessment strengthens the equilibrium, which signifies that contributing the highest quality data leads to the highest reward. In the reward layer, incentives, determined by both the quantity and quality of contributions, are granted using a reputation-weighted algorithm to ensure fair distribution. The unique Nash equilibrium in the non-cooperative data-sharing game shows that data owners act selfishly to maximize their profits.

\par Additionally, in \cite{weng2019deepchain} proposed Deepchain, which also provides reward based on the data quantity. The system involves data owners collaborating to train a model and miners processing transactions for model updates on DeepChain. Data owners pay transaction fees based on their data quantity, with miners competing to process transactions and receive rewards. Value-based incentives promote correct participant behavior. Smart contracts regulate behavior and track attackers. The system assesses global model accuracy using local updates, penalizing invalid transactions and considering updates with decreased accuracy as potentially malicious. In \cite{zhao2020privacy,wang2022Blockchain}, a customer-centric incentive system assesses contributions and calculates reputations using Multi-KRUM to eliminate unsatisfactory and malicious updates. In conjunction with this study, Abdel \textit{et al.}~\cite{abdel2022privacy} enforced a hybrid incentive strategy, incorporating Multi-KRUM for providing incentives. The authors in \cite{toyoda2019mechanism} introduce a fair and incentive-aware mechanism. Workers actively choose their top $k$ previous models during each round, assigning precisely one vote to each model. The smart contract then calculates aggregated votes, determines worker counts from the preceding round, and allocates rewards in descending order based on these counts.
\begin{table}
\centering
\caption{Privacy Preservation in BCFL using Reward Driven approaches}\label{tab:Reward}

\begin{tabular}{|p{5cm}|p{3cm}|}
\hline
\textbf{Approach} & \textbf{Reference Paper} \\
\hline
\multirow{1}{*}{Client Data Contribution} &\cite{zhao2020privacy,salim2021Blockchain,wang2022Blockchain,weng2019deepchain,hamouda2022ppss,kasyap2022efficient,qi2022high,abdel2022privacy}\\
% \cline{1-2} % Add a horizontal line
\hline

\multirow{1}{*}{Auction theory-based
schemes} &\cite{batool2022fl,kang2018Blockchain,kang2020reliable}\\
\cline{1-2} % Add a horizontal line

\multirow{1}{*}{Mechanism design-based schemes} & \cite{toyoda2019mechanism} \\
\cline{1-2} % Add a horizontal line

\multirow{1}{*}{Contract-theoretic approach} & \cite{kang2019incentive} \\
\cline{1-2} % Add a horizontal line

\multirow{1}{*}{Game theory-based schemes} & \cite{liu2023pd2s} \\
\cline{1-2} % Add a horizontal line
\multirow{1}{*}{Smart contract-based schemes} & \cite{qi2021Blockchain,qammar2023Blockchain,qu2022fl} \\
\cline{1-2} % Add a horizontal line

\end{tabular}
\end{table}

\par Rewards for edge nodes, tied to their contributions to the global model, may lack fairness and reasonability. This imbalance arises because edge nodes with substantial datasets and robust computational resources enjoy an unfair advantage, resulting in uneven reward distribution. However, \cite{qammar2023Blockchain} introduced the forward bidding mechanism, which selects the top $k$ edge nodes within the FL task publisher/server budget and compensates them accordingly. To prevent edge nodes from withdrawing during model training, they must submit a fixed security amount, refunded upon successful convergence of the global model along with the reward.

\par In certain studies, a consensus mechanism has been introduced to fairly reward legitimate users across cross-silos using the model quality. Participants earn a reputation by staking cryptocurrency deposits or their existing reputation in the Proof-of-Federated Deep-Learning~(PoFDL) consensus mechanism proposed in \cite{hamouda2022ppss}. This approach enhances trust among participants and reinforces the immutability of the Blockchain. Participants who take on the role of validator nodes gain reputation through their active involvement in the PoFDL process, establishing a mechanism where contributions to the Federated Learning system increase reputation within the network. Furthermore, Kashyap \textit{et al.}~\cite{kasyap2022efficient} introduced Proof of Interpretation and Selection~(PoIS), a consensus mechanism for participant incentives. PoIS assesses individual contributions using label-wise model interpretation through Shapley value, detecting adversaries through feature attribution aggregation. %The mechanism prioritizes honest clients with positive histories over potentially malicious high-performers. Incentives are determined by a credit function considering contribution, relevance, and past performance, ensuring fairness and security in participant rewards.
\par The authors in \cite{qu2022fl} proposed the \textquotedblleft Balanced Sign SGD \textquotedblright, a 1-bit gradient compression method that emphasizes privacy by exchanging only the signs of gradients, excluding the gradients themselves. Additionally, it introduces a novel committee-based consensus algorithm featuring a personalized incentive mechanism. It also ensures that every contributing participant is rewarded based on their distinct contributions to enhancing the model. Committee members engage in global aggregation and achieve consensus through cross-validation, with the first finisher receiving additional rewards. Other committee members are rewarded based on their response times, working as evidence of effectiveness. Participants contributing to local models receive rewards based on the cosine distance of their contributions to the global model, with rewards increasing proportionally as the cosine distance approaches predetermined thresholds. In Qi \textit{et al.}~\cite{qi2021Blockchain}, a smart contract-based reputation scheme uses the Reputation Contract~(RC) and Hunter Contract~(HC) to establish trust. The RC assigns reputation scores, rewarding positive actions and penalizing negatives. Simultaneously, the HC guards against malicious nodes by verifying weights' accuracy and reporting dishonest behavior to the RC, contributing to a trustworthy system.
\par Some studies incorporate an auction-based mechanism to reward participants efficiently, ensuring a fair and transparent compensation system for their contributions. For example, Batool \textit{et al.}~\cite{batool2022fl} proposed a multidimensional auction-based reward mechanism that utilizes a smart contract to compensate participating clients with cryptocurrencies. This auction considers factors like computational and network resources and local data quality. The reward distribution is based on the Shapley value, ensuring fairness by measuring the relative contribution of each client. Kang \textit{et al.} introduced a Subjective Logic approach, as outlined in \cite{kang2018Blockchain}, to assess individual reputations in the context of vehicular networks. This framework for probabilistic information fusion relies on subjective beliefs and operates by evaluating interactions as the basis for reputation assessment. In \cite{kang2019incentive}, the study extends \cite{kang2018Blockchain} by introducing a multi-subjective logic function to enhance the reward approach. The authors also propose a worker selection scheme for dependable Federated Learning, incorporating a multiweight subjective logic model for reputation assessment. Blockchain integration ensures secure decentralized reputation management with nonrepudiation and tamper-resistant properties. Additionally, the incentive mechanism, blending reputation and contract theory, encourages high-reputation mobile devices with quality data to engage in model learning actively. Kang \textit{et al.}~\cite{kang2020reliable} proposed Multi-weight subjective logic to enhance reputation calculation in BCFL, considering interaction attributes like frequency, timelines, and effects.
\par In \cite{liu2023pd2s}, the study proposes an incentive mechanism for a privacy-preserved data-sharing system, formulating it as a two-stage Stackelberg game. The mechanism is designed to maximize the utility of data requesters and two types of data providers, considering their distinct roles and contributions. The non-cooperative nature of the interactions justifies the choice of a Stackelberg game model, the hierarchical relationship between requesters and providers, and the one-to-many data-sharing structure. Table \ref{tab:Reward} presents an overview of privacy preservation in BCFL by enforcing a reward-driven approach using various methodologies.

%%%%%%%%%%%%%%%%%%%%%%%%%%%%%%%%%%%%%%%%%%%%%%%%%%%%%%%%%%%%%%%%%%%%
\subsection{Privacy protection using BCFL with Hybrid Privacy Approaches }
\par Several studies indicate that integrating diverse privacy approaches helps mitigate security and privacy attacks in BCFL. This section explores hybrid approaches that provide privacy by combining various privacy-preserving techniques. For instance, integrating differential privacy for initial data aggregation and applying homomorphic encryption could yield a more resilient solution. The amalgamation of HE and SMPC in BCFL markedly enhances the confidentiality and privacy of the FL process within a transparent and decentralized Blockchain framework. This integration fosters trust and security in data sharing and model training, as exemplified by \cite{feng2021Blockchain} and \cite{qu2021proof}. HE enables computations on encrypted data, preserving the privacy of individual contributions, while SMPC ensures secure collaboration among participants without exposing their raw data. Table \ref{tab:combineprivacy} summarizes privacy preservation in BCFL by employing various privacy approaches. In the privacy-focused collaborative training proposed by Zhu \textit{et al.} ~\cite{zhu2021privacy}, participants protect their local gradients using the Paillier cryptosystem with threshold decryption and a secure multi-party aggregation algorithm. This method ensures data privacy during collaborative training by transforming gradients into a secure form.

\begin{table*}
\centering
\caption{Privacy Preservation in BCFL using hybrid privacy approaches}
\renewcommand{\arraystretch}{1.4}

\begin{tabular}{|p{1.3cm}|p{3.5cm}p{2.3cm}p{4.5cm}p{2.5cm}|}
\hline
\textbf{Reference paper} & \textbf{Privacy Scheme Used} & \textbf{Parameter}& \textbf{ Attack against}&\textbf{Adversary} \\
\hline
\cite{zhu2021privacy}&HE \& SMPC& Local Gradient&I&MalC, Malicious Miners \\
\hline
\cite{zhou2022Blockchain}&HE \& SMPC& Local models&I&HbC, Clients\\
\hline
\cite{feng2021Blockchain}&HE \& SMPC&Local Models &I\&P&HbCS\\
\hline
\cite{qu2021proof}&HE \& SMPC&Local Model &I&MalC\\
\hline
\cite{hamouda2022ppss}&DP \& SMPC&Local Gradient &Byzantine and Sybil attacks, Model inversion, I, Model theft attacks&HbCS, HbCC, MalC\\
\hline
\cite{shayan2020biscotti} &DP \& SMPC&Local Gradient &I&HbC, MalC\\
\hline
\cite{jia2021Blockchain} &DP \& HE& Local weights&Model extraction attack, Model reverse attack&\\
\hline
\cite{sun2021permissioned}&DP\& HE \& SMPC& Local Gradient&Collusion attack, Sybil attack, I, \& P& HbCC, MalC\\
\hline
\cite{bai2022nttpfl}&SS, Combine Paillier and ElGamal based scheme&Local Gradient &I& Internal or External Adversary\\
\hline
\end{tabular}
\\
\vspace{0.3cm}
\footnotesize HbCS: Honest-but-Curious Server, MalC: Malicious Client, I: Inference attack, P: Poisoning attack, HbCC: Honest-but-Curious Client, SS: Secret Sharing 
\label{tab:combineprivacy}
\end{table*}

\par Furthermore, in \cite{zhou2022Blockchain} introduced a flexible and trustworthy framework for industrial intelligence, integrating autonomous FL and secure data-sharing on the Blockchain. The proposed approach preserves privacy through a combination of HE and SMPC approaches, which can enhance the security of sensitive data. Their approach involves an autonomous Federated extreme gradient boosting Learning algorithm for privacy protection, verifiability of aggregated results, and model reliability. They also introduced a secure and trusted trading mechanism for controlled on-demand data sharing, a threshold aggregation signature for model ownership assurance, and proxy re-encryption and retrieval to facilitate controllable and reliable data sharing with high accuracy and performance. Feng \textit{et al.}~\cite{feng2021Blockchain} presents a framework for decentralized cross-domain FL in 5G-enabled UAVs, leveraging Blockchain technology. It utilizes multi-signature smart contracts for dynamic cross-domain authentication, enhancing collaborative Learning. The framework employs decentralized smart contracts for model aggregation, addressing security concerns related to centralized servers. Additional security measures, such as homomorphic encryption and multiparty computation, are applied to protect against local update attacks.
\par FL presents a promising avenue for developing energy-efficient consensus algorithms, addressing the resource-intensive nature of traditional methods like PoW. Integrating the consensus process with FL eliminates the need for extra computational resources dedicated to separate consensus algorithms, potentially leading to substantial energy savings. From a communication standpoint, public Blockchains often require miners to broadcast their local model parameters, resulting in considerable communication overhead, especially as the number of miners grows. The authors in \cite{qu2021proof} proposed a method to mitigate these challenges using a novel consensus protocol like Proof-of-Federated-Learning (PoFL), leveraging the computational overhead of local training in Federated Learning as proof for consensus. PoFL significantly reduces mining power wastage and trims computational overhead while ensuring efficient consensus processes without reference to external sources.
Moreover, it proposed a novel method utilizing a reverse game-based data trading mechanism to enhance data privacy by determining optimal data trading probabilities and pricing strategies. This approach encourages data pools with high privacy risks to trade less data at a higher cost, incentivizing them to train models without data leakage. Additionally, a privacy-preserving model verification mechanism consists of HE-based label prediction and SMPC with two-party-based label comparison, ensuring model accuracy while preserving privacy for both the task requester's test data and the pool's submitted model. 
\par In \cite{hamouda2022ppss}, they explored the integration of secure multi-party computation and differential privacy to enhance system privacy. Also, a permissioned Blockchain and private peer-to-peer channels are utilized in their approach. Encourage cross-silo FL using the lightweight and energy-efficient consensus Proof-of-Federated Deep-Learning protocol, effectively detecting and classifying IIoT attacks in Non-IID and IID scenarios. Bai et al.~\cite{bai2022nttpfl} proposed a Blockchain-based privacy-preserving approach using no trusted third-party Federated Learning. They employ a conference key agreement to negotiate keys between the initiator and partners, eliminating the need for a trusted third party. A double-layer encryption mechanism ensures privacy encrypts local and global models, preventing partners from accessing each other's private information. The decentralized nature of Blockchain enhances transparency, traceability, and resilience against SPoF. Additionally, they used an efficient secret-sharing scheme to encrypt model parameters, reducing communication costs and computation time compared to Paillier and ElGamal-based schemes and secure aggregation protocols.
\par Bolstering security against Sybil attacks, poisoning attacks, and inference attacks, Shayan \text{et al.}~\cite{shayan2020biscotti} incorporate differential privacy and encryption approach within BCFL with secure and private multi-party ML. In each iteration, peers compute local model updates, keeping them private by masking with differentially private noise obtained from a set of peers identified through a verifiable random function. Verification committees validate these masked updates to prevent poisoning. If the majority of the committee approves, the updates are divided into Shamir's secret shares and passed to an aggregation committee. This committee securely aggregates the unmasked updates, with contributing peers and committee members receiving additional stake in the system. The aggregated updates are then added to the global model within a newly created Blockchain block and shared with all peers, and the process repeats with the updated global model and stake.
\par The studies outlined in \cite{jia2021Blockchain} focus on establishing a secure data-sharing mechanism to uphold privacy among numerous distributed users. It also suggests a data protection aggregation approach that utilizes distributed K-means clustering with DP and HE, random forest with DP, and AdaBoost with HE to enhance data protection in Industrial IoT scenarios. Sun \textit{et al.}~\cite{sun2021permissioned} address the challenge of enhancing security and privacy in their work. They use a Blockchain to record each global model update, ensuring the verifiability and traceability of local updates through permanent records. It also enables an incentive mechanism tailored to user contributions.
Additionally, HE secures users' local model updates. A validation process precedes local update aggregation to thwart poisoning attacks, and privacy is maintained with differential privacy noise. Ultimately, they establish a secure aggregation scheme for local updates using the Shamir secret sharing technique, balancing utility and privacy compared to differential privacy.

Table \ref{tab:Blockchain-fl-summary} elucidates the overview of studies specifically in BCFL, highlighting their privacy approach, Blockchain types,  the Blockchain frameworks utilized within Federated Learning systems, consensus algorithms, and block storage techniques.

%%%%%%%%%%%%%%%%%%%%%%%%%%%%%%%%%%%%%%%%%%%%%%%%%%%%%%%%%%%%%%%%%%%%
\begin{table*}
\centering
\caption{Summary of studies on integration of Blockchain enabled Federated Learning, elucidating their privacy preservation methods, types of Blockchain used, the Blockchain frameworks integrated within Federated Learning systems, consensus algorithms employed, block storage, and data distribution used.}\label{tab:Blockchain-fl-summary}

\begin{tabular}{|c|p{4cm}p{5.5cm}|}
\hline
    & \textbf{Techniques} & \textbf{Reference Paper} \\
\hline
\multirow{11}{*}{\textit{Privacy Approach}} & Differential privacy & \cite{zhao2020privacy,lu2020Blockchain,xu2023Blockchain,lu2019Blockchain,qu2020decentralized,li2021byzantine,qiy2021privacy,wang2020Learning,wan2022privacy,qu2022fedtwin,cui2021security,zhang2021Blockchain,javed2023sharechain,liu2023pd2s,li2021Blockchain,ji2023lafed,salim2021Blockchain, zhang2023privacy,chen2023privacy,arachchige2020trustworthy}\\
\cline{2-3} % Add a horizontal line
& Homomorphic encryption &  \cite{qi2021Blockchain,li2023privacy,guduri2023Blockchain,chen2021ds2pm,wang2022Blockchain,weng2019deepchain,sun2022Blockchain,sezer2023ppfchain,muazu2023iomt,li2021privacy,qu2021proof,chen2022esb}\\
\cline{2-3} % Add a horizontal line
 & Secure multi-party computation & \cite{awan2019poster,li2020Blockchain,jiang2021pflm,fang2022privacy,wang2020ai}\\
 \cline{2-3} % Add a horizontal line
 & Reward driven approaches & \cite{kang2019incentive,xu2023Blockchain,singh2022framework,lu2020low,feng2021bafl,guo2022sandbox,wang2020Learning,liu2023pd2s,li2021Blockchain,salim2021Blockchain,miao2022intelligent,wang2022Blockchain,weng2019deepchain,qi2022high,abdel2022privacy,batool2022fl,kang2018Blockchain,hamouda2022ppss,kang2020reliable}\\
 \cline{2-3} % Add a horizontal line
& Hybrid privacy approaches &  \cite{feng2021Blockchain, qu2021proof,hamouda2022ppss,zhu2021privacy,jia2021Blockchain,zhou2022Blockchain,bai2022nttpfl,shayan2020biscotti,sun2021permissioned} 
\\
\hline
\multirow{13}{*}{\textit{Consensus Protocol}} & PoW &\cite{qu2020decentralized,singh2022framework,kumar2021Blockchain,feng2021bafl,wang2020Learning,wan2022privacy,cui2021security,li2021Blockchain,salim2021Blockchain,muazu2023iomt,li2021privacy,wang2020ai}\\
\cline{2-3} % Add a horizontal line
& PoS &  \cite{chen2022esb} \\
\cline{2-3} % Add a horizontal line
 & DPoS & \cite{lu2020Blockchain,lu2020low} \\
 \cline{2-3} % Add a horizontal line
 & pBFT &\cite{kang2019incentive,otoum2020Blockchain,qi2021Blockchain,guo2022sandbox,yang2022trustworthy,qiy2021privacy,liu2023pd2s,wang2022Blockchain,feng2021Blockchain,sun2021permissioned,kang2020reliable}\\
 \cline{2-3} % Add a horizontal line
& PoA & \cite{li2021byzantine,javed2023sharechain}\\
\cline{2-3} % Add a horizontal line
& PoQ & \cite{lu2019Blockchain}\\
\cline{2-3} % Add a horizontal line

& PoF & \cite{li2021byzantine,qu2022fedtwin,qu2021proof,shayan2020biscotti}\\
\cline{2-3} % Add a horizontal line \\
& PoC &\cite{sezer2023ppfchain}\\
\cline{2-3} % Add a horizontal line

& PoFL &  \cite{qu2021proof}   \\
\cline{2-3} % Add a horizontal line
& RAFT&  \cite{feng2021Blockchain,jia2021Blockchain} 
\\
\cline{2-3} % Add a horizontal line
& Con-dBFT &  \cite{chen2021ds2pm}  \\

\cline{2-3} % Add a horizontal line
& Algorand & \cite{zhao2020privacy,weng2019deepchain,abdel2022privacy}  \\
\hline
%Blockchain Type
\multirow{7}{*}{\textit{Blockchain Type}} &  Public &\cite{lu2020Blockchain,otoum2020Blockchain,kumar2021Blockchain,guduri2023Blockchain,li2021byzantine,wan2022privacy,qu2022fedtwin,cui2021security, javed2023sharechain,arachchige2020trustworthy,abdel2022privacy,batool2022fl,jia2021Blockchain}\\
\cline{2-3} % Add a horizontal line
& Private & \cite{miao2022privacy,kalapaaking2023Blockchain} 
\\
\cline{2-3} % Add a horizontal line
& Permissioned &  \cite{li2023privacy,lu2020low,kasyap2023privacy,sarhan2022hbfl,zhou2022Blockchain,hamouda2022ppss}\\    
\cline{2-3} % Add a horizontal line
& Consortium & \cite{zhao2020privacy,kang2019incentive,qi2021Blockchain,qu2020decentralized,xu2022efficient,guo2022sandbox,qiy2021privacy,wang2020Learning,qu2022fedtwin,liu2023pd2s,chen2021ds2pm,wang2022Blockchain,sun2022Blockchain,muazu2023iomt,qi2022high,kang2018Blockchain,kang2020reliable,feng2021Blockchain,bai2022nttpfl}\\
\hline
%Blockchain Platform
\multirow{5}{*}{\textit{Blockchain Platform}} & Ethereum &\cite{awan2019poster,qammar2023Blockchain,li2023privacy,guduri2023Blockchain,javed2023sharechain,miao2022intelligent,arachchige2020trustworthy,miao2022privacy,muazu2023iomt,zhou2022Blockchain,jiang2021pflm,abdel2022privacy,batool2022fl,zhu2021privacy}\\
\cline{2-3} % Add a horizontal line
& Hyperledger Fabric &  \cite{qu2020decentralized,kasyap2023privacy,xu2022efficient,guo2022sandbox,wang2022Blockchain,sun2022Blockchain,qi2022high,feng2021Blockchain,sun2021permissioned}\\
\cline{2-3} % Add a horizontal line

& Custom Blockchain &  \cite{salim2021Blockchain} \\
\hline
% Blockchain storage
\multirow{5}{*}{\textit{Blockchain Storage}} & off-chain &\cite{zhao2020privacy,otoum2020Blockchain,awan2019poster,qi2021Blockchain,qammar2023Blockchain,qu2020decentralized,guduri2023Blockchain,li2021byzantine,zhang2021Blockchain,javed2023sharechain,liu2023pd2s,ji2023lafed,chen2023privacy,arachchige2020trustworthy,chen2021ds2pm,sezer2023ppfchain,batool2022fl,feng2021Blockchain,shayan2020biscotti}\\
\cline{2-3} % Add a horizontal line
& on-chain & \cite{li2023privacy,xu2022efficient,feng2021bafl,guo2022sandbox,wan2022privacy,cui2021security,liu2023pd2s,zhang2023privacy,sun2022Blockchain,sezer2023ppfchain,muazu2023iomt}\\
% \cline{2-3} % Add a horizontal line
% & IPFS &   \\

\hline
% Data Distribution
\multirow{7}{*}{\textit{Data Distribution}} & IID &\cite{zhao2020privacy,lu2020Blockchain,xu2023Blockchain,otoum2020Blockchain,chen2022esb,qu2020decentralized,lu2020low,guduri2023Blockchain,li2021byzantine,xu2022efficient,feng2021bafl,yang2022trustworthy,wan2022privacy,cui2021security,liu2023pd2s,ji2023lafed,salim2021Blockchain,chen2023privacy,arachchige2020trustworthy,weng2019deepchain,chen2021ds2pm,sun2022Blockchain,miao2022privacy,kalapaaking2023Blockchain,qi2022high,abdel2022privacy,hamouda2022ppss,qu2021proof,shayan2020biscotti}
\\
\cline{2-3} % Add a horizontal line
& Non-IID&  \cite{kasyap2023privacy,guo2022sandbox,yang2022trustworthy,qu2022fedtwin,li2021Blockchain,miao2022intelligent,wang2022Blockchain,qi2022high,abdel2022privacy,hamouda2022ppss,kasyap2022efficient,zhu2021privacy}\\
% \cline{2-3} % Add a horizontal line
\hline

\end{tabular}

\end{table*}

%%%%%%%%%%%%%%%%%%%%%%%%%%%%%%%%%%%%%%%%%%%%%%%%%%%%%%%%%%%%%%%%%%%%

%%%%%%%%%%%%%%%%%%%%%%%%%%%%%%%%%%%%%%%%%%%%%%%%%%%%%%%%%%%%%%%%%%%%

\section{Privacy Preservation using Cross-chained FL Approaches}
\label{sec:solution2}
In this section, we have delved into the intricacies of cross-chain-enabled Federated Learning as a mechanism for preserving privacy. The discussion thoroughly explores how leveraging cross-chain capabilities enhances Federated Learning methodologies to uphold and safeguard privacy.
\subsection{Overview of Cross-chained FL}
Recent studies indicate that BCFL systems preserve the system's privacy. Still, the limited scalability of a single Blockchain becomes evident as the number of FL training tasks increases, resulting in the simultaneous generation of numerous blocks and subsequent queuing for block verification. This scalability challenge emerges due to the difficulty of managing massive block data with a limited number of miners, leading to constrained throughput, reduced efficiency, and slower FL training processes~\cite{beltran2023decentralized}. Additionally, BCFL incurs a substantial communication cost for model update transmission, requiring multiple rounds of communication to achieve the desired accuracy level. This arises from frequent gradient exchanges among peers over limited bandwidth channels, and as the block data size increases, so does the flow of model updates across the Blockchain network, posing significant communication challenges.
Moreover, Blockchain-enabled Federated Learning encounters numerous challenges, including selecting efficient miners, consensus algorithm implementation, and chain validation \cite{imteaj2021survey,majeed2019flchain,li2020Blockchain}. Cross-chain technology enables data exchange among multiple Blockchains. Which also facilitates secure data transfers while maintaining the same machine-learning models throughout various Blockchain networks~\cite{jiang2020Blockchain,qu2022Blockchain}. The following highlights the major benefits and key advantages of cross-chained enabled FL~\cite{kang2022communication}.
\begin{itemize}
    \item \textbf{Higher Scalability:} Cross-chained FL outperforms single Blockchain in efficiency and scalability. Unlike single Blockchain limitations in managing FL training tasks, cross-chained systems efficiently distribute workloads, mitigating bottlenecks. The parallel processing capability ensures optimal scalability, seamlessly accommodating growing FL task demands. Multiple interconnected Blockchains enhance resource management, improving system efficiency and security compared to a singular Blockchain system. The cross-chain integration in FL enables global collaboration, fostering diverse participation and data federation across regions and industries. 
    \item \textbf{Low Communication Cost:} Blockchain-based FL requires frequent gradient exchanges to synchronize model updates among peers, utilizing limited bandwidth channels. Cross-chained FL networks employ a compressed gradient strategy, ensuring cost-effectiveness and high accuracy. Due to the compression of gradients, this scheme fortifies the safeguarding of training data privacy by reducing the efficacy of gradient leakage attacks when there is an inadequate amount of gradient information~\cite{zhu2019deep}.%Their enhanced privacy protection mechanisms give them an advantage over single Blockchains in addressing communication cost challenges.
    \item \textbf{Reduced Single-Point-of-Failure Risks:} Cross-chain Federated Learning mitigates the risks associated with a single-point-of-failure. Distributing the learning process across multiple Blockchain's makes the system more resilient to potential disruptions or attacks on a single chain.
\end{itemize}

\subsection{Solutions for privacy preservation using cross-chain approaches}

\par In this section, we explored the intricacies of the cross-chained network, which has diverse privacy solutions meticulously crafted to safeguard the system's privacy by integrating cross-chain-enabled Federated Learning. Kang \textit{et al.}~\cite{kang2022communication} introduced an innovative cross-chain powered FL framework with parallel Blockchains designed to handle model updates securely, with scalability and flexibility, eliminating the constraints of conventional single BCFL systems. Their approach incorporated a two-phase commit protocol to validate and authenticate block data across multiple Blockchains for Artificial Intelligence of Things in 6G. Furthermore, they utilized a mixed-precision local training strategy combined with flexible model update compression to improve communication efficiency without compromising accuracy. In the \textquotedblleft Prepare\textquotedblright phase, the system establishes the groundwork by deploying model training and payment smart contracts on the source and destination parachains. The task publisher calls the training smart contract, sends a cross-chain request, and collaborates with validators and collators across parachains for legitimacy. Simultaneously, the payment smart contract activates to secure assets for worker rewards upon model training completion. Transitioning to the \textquotedblleft Commit\textquotedblright phase, the trained model undergoes quality evaluation, triggering the training smart contract to generate a Simplified Payment Verification~(SPV) proof and block header. Verified by the relay chain's validator group, they reach a consensus on the model training's legitimacy. Successful validation leads to worker compensation, with payment records logged in the payment chain. Discrepancies prompt a rollback, releasing locked assets. This two-phase process ensures the secure execution of cross-chain-enabled Federated Learning, managing complexities across interconnected Blockchains.
\par The prevailing BCFL system encounters data sparsity issues despite its commendable system efficiency. To tackle these concerns, Jin \textit{et al.}~\cite{jin2021Cross} introduced a cross-cluster Blockchain-enabled FL framework employing a cross-chain approach for the Internet of Medical Things. Their proposal includes the integration of two Blockchain consensus algorithms to facilitate secure model exchange across clusters using PBFT and a two-phase cross-chain consensus mechanism.
Additionally, they advocate for model aggregation within each BCFL cluster and subsequent transmission to the other cluster, resulting in a remarkable enhancement of system efficiency and accuracy, with performance increased from 39.3\% to 75.8\%. This places a significant burden on computational and communication resources, so researchers suggested using it with edge computing instead of end devices. Kang \textit{et al.}~\cite{kang2023Blockchain} introduced a privacy framework that employed a hierarchical cross-chain structure for healthcare metaverses. The proposed system empowers users to safeguard sensitive data in the physical space and contribute non-sensitive data for metaverse tasks. Also, a data freshness-based incentive mechanism inspired by prospect theory~\cite{kahneman2013prospect} is used for user-centric data sharing, and a pallier homomorphic encryption algorithm is used to provide security and privacy. Their approach achieved 93.71\% accuracy in breast cancer prediction via vertical FL training.
\par Xu \textit{et al.}~\cite{xu2022mudfl} present a hierarchical micro chained fabric, denoted as µDFL, designed for decentralized, Federated Learning across devices in edge networks. The microchain consensus protocol, built upon a partially decentralized Blockchain utilizing Proof-of-Credit~(PoC), ensures the transparency and privacy of data sharing during local model training. The proposed µDFL introduces a hierarchical Internet of Things network fabric, incorporating lightweight microchains. Each microchain adopts a hybrid approach involving PoC block generation and a Voting-based Chain Finality consensus to enhance efficiency and privacy. The Federated structure of µDFL is achieved through an inter-chain network employing Byzantine Fault Tolerance. Validation through a proof-of-concept prototype demonstrates the effectiveness of µDFL in cross-device Federated Learning environments, emphasizing efficiency, security, and privacy. 

\section{Application of Blockchain-enabled FL for Privacy Preservation}
\label{sec:application}

In the following sections, we describe how several approaches leveraging FL and Blockchain for privacy preservation are used in different application scenarios, such as Healthcare, Industrial IoT (IIoT), and the Internet of Vehicles.

\subsection{Healthcare}

The analysis of health data using ML techniques can result in therapies and procedures with lower risks and better outcomes for patients, thus increasing the quality of care \cite{antunes2022Federated}.
Healthcare data are usually spread across
various sources such as hospitals, clinics, and wearable devices, which are characterized by highly sensitive information demanding to keep patients' data as private as possible. The decentralized nature of Blockchain technology and the ability of FL solutions to train models locally, while sharing only model parameters, has made the combination of the two approaches well-suited for healthcare.
Moreover, Blockchain-based FL not only overcomes challenges associated with the outflow of confidential medical data efficiently but can {\em(i)} reward FL members for their contribution to the network {\em(ii)} monitor that the centralized FL server accurately aggregates the global model.

In this context, multiple IoT devices, including weight meters, blood pressure, glucose meters, insulin pumps, and others are connected to patients and aim at acquiring specific data they are meant to be gathered from the human body, such as temperature, heartbeat, electrocardiograph, and many others.
These devices communicate data to smart systems such as smart monitors, laptops, and mobiles to be analyzed and visualized. The main goals of the different Blockchain-based FL approaches for healthcare can be summarized as follows \cite{myrzashova2023Blockchain}:

\begin{itemize}
    \item management of medical records, also thanks to the cooperation between multiple hospitals/systems;
    \item tracking disease outbreak;
    \item enhanced monitoring of patients thanks to a wider amount of data to be analyzed;
    \item improving sensors' performance;
    \item pharmaceutical clinical trials.
\end{itemize}

In the system presented in \cite{singh2022framework}, IoT devices, before communicating with third-party components, send data to a Blockchain network for validation and, after this step, data is forwarded to other systems. Moreover, Blockchain provides large independent storage for healthcare data, recording usage behavior and ensuring authenticity. Multiple actors collaborate to provide a privacy-preserving solution, namely: {\em(i)} the sub-feature manager, which vertically partitions the aggregated data into different datasets; {\em(ii)} the different clients, which provide data to the federation manager and receives a sub-model for the training; {\em(iii)} the privacy broker, in charge of solving privacy issues; {\em(iv)} the integrity manager, which maintains the result integrity by avoiding errors inside sub-models.

The authors in \cite{kalapaaking2023Blockchain} proposed Blockchain architecture assumes the presence of numerous smart hospitals situated in diverse regions, each equipped with a cluster of IoT medical devices. These devices use edge servers for FL tasks with privacy-preserving verification via SMPC before aggregation. After verification, the local model is sent to the Blockchain for SMPC-based secure aggregation. Once a consensus is reached, the global model is stored in the Blockchain, and tamper-proof storage shares it with all FL round hospitals. 

Also, the works presented in \cite{lakhan2022Federated,polap2021agent}
are related to the Internet of Medical Things~(IoMT) devices. In particular, \cite{lakhan2022Federated} proposes a Blockchain-enabled Federated Learning in the context of IoMT, with privacy-preservation and fraud detection characteristics. The solution is intended for healthcare applications in a fog-cloud-assisted network. The authors of \cite{polap2021agent} introduce a real-time medical data processing multi-agent system that utilizes Blockchain for sharing and safeguarding private data.

Similarly to the previous approach, the architecture presented in \cite{kumar2021Blockchain} considers multiple hospitals leveraging FL to keep their data private, thus sharing only weights and gradients. In this case, the aim is to recognize the presence of COVID-19 infection from lung Computed Tomography (CT, hereafter) scans. 
Each hospitals use Blockchain technology to distribute data, with each hospital storing an actual CT scan, and Blockchain facilitating the retrieval of the trained model. Privacy is ensured through encryption and the storage of unique identifiers for each hospital.

In the realm of COVID-19 diagnosis, \cite{samuel2022iomt} introduces a Federated Blockchain-powered medical system, termed FedMedChain. The primary objective of this system is to distribute COVID-19 information and establish a collaborative diagnosis model while safeguarding the privacy of data owners.

In the paper presented in \cite{hai2022bvflemr}, the authors propose a framework called Blockchain Vertical Federated Learning E-Medical Recommendation~(BVFLEMR). It adopts a decentralized digital ledger system for Electronic Health Records~(EHR) storage, LightGBM, and N-Gram models to recommend tailored treatments for the patients based on their EHR. In this way, it achieves private storage and management of patients' sensitive data in EHRs, such as diagnosis, treatment, medication, surgery, and diet specifications.

Privacy of EHRs is taken into account also by the authors of \cite{alzubi2022cloud}, who propose a framework called CNN\_BC\_Cryp\_FL. It consists of {\em(i)}, a CNN-based secure classification component able to classify normal and abnormal users using the available dataset; 
{\em(ii)} a Blockchain-integrated cryptography-based FL used to restrict the accessibility of the database to abnormal users.

Several studies leverage Blockchain to incentivize participants to contribute their local data in training FL tasks \cite{chen2022esb,liu2022Blockchain,lo2022toward}.
In particular, the work presented in \cite{chen2022esb} describes the system called ESB-FL that can train a model while protecting
the privacy of local training data using a function encryption scheme called non-interactive designated decryptor function encryption (NDD-FE). It also integrates Blockchain to support
the fair payment between the task publisher and all
participants, thus guaranteeing that each participant gets a reward if the trained model
satisfies the task requirements. Instead, the authors of \cite{lo2022toward} aim to improve the fairness of the federated learned model and the trustworthiness of medical diagnostic image analyses to detect COVID-19. 

Table \ref{tab:healthcareApp} shows the main differences among the papers analyzed in this section in terms of goal, type of involved devices, and type of healthcare data.
\begin{table*}
\centering
\caption{Healthcare Applications}
\renewcommand{\arraystretch}{1.4}
\begin{tabular}{|p{2.5cm}|p{5cm}p{3.6cm}p{3cm}|}
\hline
\multicolumn{1}{|c|}{\textbf{Reference paper}} & \textbf{Aim} & \textbf{Type of devices generating data} & \textbf{Type of Data}\\
\hline
Singh et al.\cite{singh2022framework} &Private storage, Health alert& IoT sensors~(weight meters, blood pressure, glucose meter, insulin pump) & Temperature, heartbeat, blood pressure, electrocardiograph \\
\hline
Kalapaaking et al.\cite{kalapaaking2023Blockchain}, Polap et al. \cite{polap2021agent}& Privacy-preserving analysis from multiple hospitals&Internet of Medical Things devices& Medical datasets\\
\hline
Lakhan et al.\cite{lakhan2022Federated} & Fraud analysis, and Data validation&Internet of Medical Things (IoMT) devices& Medical datasets\\
\hline
Kumar et al.\cite{kumar2021Blockchain} & Diagnosis of COVID-19& CT device&  Lung Computed Tomography scans\\
\hline
Samuel et al.\cite{samuel2022iomt} & 
Privacy-preserving Diagnosis and Dissemination of COVID-19 &Internet of Medical Things (IoMT) devices& Medical datasets\\
\hline
Hai et al.\cite{hai2022bvflemr} & EHR Private storage, Recommendation for tailored treatment& Manual data insertion & Electronic Health Records (EHR)\\
\hline
Alzubi er al. \cite{alzubi2022cloud} & Abnormal users identification and Database access& Manual data insertion & Electronic Health Records (EHR)\\
\hline
Chen et al. \cite{chen2022esb} & Privacy-preserving image detection, and incentive mechanism& Manual data insertion & Chest X-ray images\\
\hline
Liu et al. \cite{liu2022Blockchain} & Privacy-preserving image detection, and incentive mechanism & Manual data insertion & Skin Cancer images\\
\hline
Lo et al. \cite{lo2022toward} & Diagnosis of COVID-19, and incentive mechanism & Manual data insertion &  X-rays images\\
\hline
\end{tabular}
\label{tab:healthcareApp}
\end{table*}

\vspace{0.2cm}
\subsection{Industry 5.0}

Industry 5.0 is a new concept that focuses on the cooperation between humans and machines to create sustainable industrial products and services.
The main principles inspiring this innovative scenario, (namely, sustainability, human-centeredness, and resilience) are obtained thanks to the integration of digital technologies, the Industrial Internet of Things (IIoT, hereafter), artificial intelligence, and other advanced technologies into the manufacturing and industrial processes \cite{leng2022industry}.
In this context, the combination of FL and Blockchain could provide powerful solutions for industries seeking to leverage data for innovation, while ensuring privacy, security, and efficiency.

In \cite{hamouda2022ppss}, the authors present a framework called PPSS to protect privacy and defend against cyber attacks in the context of industry
4.0/5.0. PPSS includes two modules: {\em(i)} a Blockchain-enabled FL scheme, leveraging a differentially private training strategy, with an energy-efficient consensus protocol, named Proof-of-Federated Deep-Learning (PoFDL), and {\em (ii)} a privacy-preserving intrusion detection scheme using Convolutional Neural Networks for attack identification.

Similarly, \cite{yazdinejad2022block} proposes a Federated threat-hunting model
in IIoT networks to identify anomalous behavior, while
preserving the privacy of IIoT devices related to Blockchain-based smart factories.

The goal of the works proposed in \cite{lu2019Blockchain,chen2023privacy,jia2021Blockchain} is to design a secure data-sharing mechanism, that can share data among multiple distributed users while maintaining data privacy. In particular, the paper presented in \cite{lu2019Blockchain} integrates FL in the consensus process of a permissioned
Blockchain, so that the computing work for consensus can
also be used for Federated training tasks. Whereas, \cite{chen2023privacy} uses FL to obtain privacy-preserving model training, the InterPlanetary File System~(IPFS) distributed storage system for storing model parameters and generating
corresponding addresses based on the content, and Blockchain to provide the provenance and immutability of the parameters.
Instead, the work of \cite{jia2021Blockchain} proposes a data protection aggregation scheme based
on three ML methods (i.e., distributed K-means clustering based on differential privacy and homomorphic encryption, distributed random forest with differential privacy, and distributed AdaBoost with homomorphic encryption) to enable multiple data protection in IIoT scenarios.

Always in the context of secure data sharing, the paper described in \cite{yang2022privacy} tackles the problem of privacy-preserved credit data sharing. 
The combined credit data storage mechanism with a Deletable Bloom filter (DLBF) guarantees the traceability of the entire credit data sharing
process in industrial applications. Moreover, they leverage homomorphic encryption, FL, and Blockchain to avoid data leakage. 
The paper presented in \cite{zhang2020Blockchain}
has a different goal. It focuses on preserving the privacy of the client data (e.g., usage frequency and time) adopting FL to train the models locally on the client to detect possible device failures in the network. Moreover, to resolve
disputes between the central organization and client organizations about failure causes, the architecture leverages a combination of Blockchain and Merkle-tree to enable verifiable integrity of client data. 

The authors of \cite{singh2023fusionfedblock}
develop a framework for FL tasks to preserve privacy among various industrial departments. Decentralized secure storage is
provided by the Distributed Hash Table (DHT) at the cloud layer of the proposed scheme, while the Blockchain network provides data authentication and validation.

Industry 5.0 is also expected to reshape the agriculture industry, as already done in the past, and promote the fourth agricultural revolution \cite{liu2020industry}.
In this context, the authors of \cite{friha2022felids}
propose an intrusion detection system, called FELIDS, for
securing agricultural-IoT infrastructures. It aims to protect data privacy through FL, employing three deep Learning classifiers, namely, Deep Neural
Networks, Convolutional Neural Networks, and Recurrent Neural Networks against Agricultural IoT attacks. Moreover, Blockchain helps network members to track relevant information for supply chain management.

Table \ref{tab:IndustryApp} summarizes the main aim of the papers analyzed in this section.

\begin{table}
\centering
\caption{Industry 5.0 applications}
\renewcommand{\arraystretch}{1.4}

\begin{tabular}{|p{3.9cm}|l|}
\hline
\multicolumn{1}{|c|}{\textbf{Reference paper}} & \textbf{Main Aim} \\
\hline
Hamouda et al.\cite{hamouda2022ppss} & Privacy-preserving FL, and Intrusion Detection  \\
\hline
Yazdinejad et al.\cite{yazdinejad2022block}, Friha et al. \cite{friha2022felids} & Privacy-preserving Anomalies Detection \\
\hline
Lu et al.\cite{lu2019Blockchain}, Chen et al.\cite{chen2023privacy}, Jia et al.\cite{jia2021Blockchain}  & Privacy-preserving Data Sharing\\
\hline
Yang et al. \cite{yang2022privacy} &  Privacy-preserving Credit Data Sharing \\
\hline
Zang et al.\cite{zhang2020Blockchain} & Privacy-preserving Device Failure Detection\\
\hline
Singh et al.\cite{singh2023fusionfedblock} & Privacy-preserving FL \\

\hline
\end{tabular}
\label{tab:IndustryApp}
\end{table}

\subsection{Internet of Vehicles}

Internet of Vehicles (IoV, hereafter) defines the evolution of the conventional Vehicle Ad-hoc Networks and enables real-time information exchange among all the actors traveling through streets (e.g., vehicles, drivers, pedestrians) and road infrastructure through vehicle-to-everything (V2X) communication. The objective of IoV is to realize the convergence of mobile communication technology, intelligent transportation, and information systems \cite{yang2014overview,kaiwartya2016internet}.
Because this scenario allows for quick and efficient exchange of large amounts of data containing private information (i.e. location and user preferences), approaches that rely on the combination of FL and Blockchain have been investigated. The arisen challenges are the following:

\begin{itemize}
    \item strengthening privacy protection mechanisms;
    \item preventing hostile intelligent connected vehicles (ICVs) and edge servers from faking FL aggregate results with verification mechanisms;
    \item reducing the high communication overhead of FL. 
\end{itemize}

The papers proposed in \cite{smahi2023bv,lu2020Blockchain,wang2022Blockchain} provide approaches for data sharing among
vehicles for collaborative analysis to enhance service quality and driving experience.
In particular, in \cite{smahi2023bv}, a Blockchain-enabled and privacy-preserving FL framework called BV-ICVs is presented. In this system, smart contracts are used to prevent malicious ICVs from uploading unreliable, erroneous, or low-quality FL model updates. The authors of \cite{lu2020Blockchain} use FL to relieve transmission load and address privacy concerns of providers, and a hybrid Blockchain architecture, which consists of a permissioned Blockchain and a local Directed Acyclic Graph (DAG), executing a two-stage verification to obtain the reliability of shared data. The scheme illustrated in \cite{wang2022Blockchain}, in addition to a Blockchain-based Privacy-preserving Federated Learning approach, also proposes a reputation-based model as an incentive mechanism to encourage users of IoV to participate in FL tasks actively.

The work presented in \cite{qiy2021privacy} describes a framework for traffic flow prediction. To avoid using a centralized model coordinator, a consortium Blockchain-based FL framework is proposed to enable decentralized and secure FL. The model updates from distributed vehicles are verified by miners and stored on the Blockchain. Moreover, to preserve model privacy on the Blockchain, a differential privacy method with a noise-adding mechanism is used. Likewise, the system proposed in \cite{pokhrel2020Federated}
aims at removing the FL centralized global server and using a Blockchain to exchange local model updates from vehicles while providing and verifying their corresponding rewards.

The authors of \cite{chai2020hierarchical} focus on an approach to provide a hierarchical Blockchain-enabled FL algorithm for knowledge sharing in IoV. Moreover, they formulate a lightweight Proof-of-Knowledge (PoK) consensus mechanism to reduce the computation consumption. The works presented in \cite{liu2021Blockchain,moulahi2023privacy} have the different goal of designing a cooperative intrusion detection mechanism that offloads the training model to IoV devices. \cite{liu2021Blockchain} distributes the FL computation to reduce the resource utilization of a central server while assuring security and privacy, and it relies on Blockchain to ensure the
security of the aggregation model, store, and share the training models. Instead, \cite{moulahi2023privacy} uses Blockchain to store and share models from the previous steps in a smart contract and return the updated models to the vehicles.

An IoV-related application scenario is that of Drone Edge Intelligence, which refers to the ability of unmanned aerial vehicles (UAVs), or drones, to process and analyze data directly at the source or edge rather than relying on a centralized computing system. Drones' characteristics, such as line of sight, ease of deployment, and capture of high-resolution images, make them the efficient solution for disaster mitigation, security surveillance, environmental monitoring, and recovery \cite{alsamhi2021multi}. FL allows drones to execute decentralized collaborative learning by computing local models. Only model parameters are shared with neighbors and the centralized unit to improve global model accuracy, while still keeping local data private. On the other hand,
Blockchain can enable privacy-preserving data sharing in a distributed manner. However, combining the two solutions raises several challenges, such as scalability, energy efficiency, and transaction capacity \cite{alsamhi2021drones}. The paper presented in \cite{akram2023chained} relies on Blockchain technology and FL for privacy-preserving malicious node detection in the Internet of Drone Things
(IoDTs).

Table \ref{tab:IoVApp} summarizes the main aim of the paper analyzed in this section.

\begin{table}
\centering
\caption{Internet of Vehicles applications}
\renewcommand{\arraystretch}{1.4}

\begin{tabular}{|p{3.9cm}|l|}
\hline
\textbf{Reference paper} & \textbf{Main Aim} \\
\hline
Smahi et al.\cite{smahi2023bv}, Lu et al.~\cite{lu2020Blockchain}, Wang et al.~\cite{wang2022Blockchain} & Privacy-preserving and verifiable FL\\
\hline
Qi et al.\cite{qiy2021privacy} & Privacy-preserving traffic flow prediction\\
\hline
Pokhrel et al.\cite{pokhrel2020Federated} & Privacy-preserving distributed FL \\
\hline
Chai et al.\cite{chai2020hierarchical}& Privacy-preserving knowledge sharing\\
\hline
Liu et al.\cite{liu2021Blockchain}, Moulahi et al.\cite{moulahi2023privacy}& Privacy-preserving and cooperative intrusion detection\\
\hline
Akram et al.\cite{akram2023chained}& Privacy-preserving malicious drone detection \\

\hline
\end{tabular}
\label{tab:IoVApp}
\end{table}

\section{Open Issues and Future Directions}
\label{sec:future}
Integrating Blockchain technology into Federated Learning is a noteworthy research area, offering substantial improvements in protecting privacy models. As highlighted earlier, BCFL is crucial in supporting various domain applications. In this section, we address issues in BCFL and propose potential solutions to shed light on future research directions for readers and researchers in this evolving domain.
\subsection{Open Issues}
\begin{itemize}

    \item \textbf{Privacy Issues in BCFL: }Preserving data privacy in BCFL involves a delicate balance between cryptographic tools and lightweight techniques. While private aggregation using cryptographic tools provides robust parameter privacy, it is computationally intensive and limited in arithmetic operations. On the other hand, noise perturbation methods, such as adding noise and gradient compression, offer a more lightweight approach but come with a trade-off between model performance and data privacy. %The predominant focus in current research involves using HE to protect against inference attacks. However, a significant drawback of HE is its low computational efficiency and inability to handle large and complex operations, posing challenges in ensuring privacy for extensive and intricate datasets.
    \par Current research predominantly emphasizes using HE to safeguard against inference attacks. However, a noteworthy drawback of HE is its limited computational efficiency and inability to handle large and complex operations efficiently. This limitation presents challenges in ensuring privacy for extensive and intricate datasets. Additionally, it is essential to note that HE does not inherently guard against collusion attacks, and this vulnerability remains a concern even when utilizing HE for privacy protection in data-intensive scenarios~\cite{hao2019efficient}.
    \par Integrating SMPC in BCFL presents notable issues. Firstly, SMPC's interactive nature clashes with the noninteractive protocol required for secure aggregation in BCFL. Additionally, the susceptibility to collusion attacks poses a significant threat even when employing SMPC to protect the model, undermining the overall security and integrity of the FL process within the Blockchain environment. Addressing these issues is crucial for fortifying BCFL against collaboration threats and maintaining trust in the collaborative model training process. In BCFL, employing HE or SMC-based methods may prove impractical for large-scale scenarios due to the considerable increase in communication and computation expenses.
    \item \textbf{Computation Overheads in BCFL:} Researchers' incorporation of encryption methods aims to bolster privacy but comes with computational overhead as encrypted gradients are transmitted on the Blockchain. The size of local model gradients plays a role in determining the communication overhead. In addressing this, some studies employ gradient compression methods to reduce overhead, though this introduces potential issues such as removing pertinent information during compression. These compression-related challenges may have repercussions on the performance of the global model~\cite{cui2020creat}.
    \item \textbf{Gas consumption in HE based system:} In \cite{qi2021Blockchain}, an incentive mechanism has been introduced to the BCFL system, leveraging HE applied to local gradients to simultaneously provide rewards and preserve participant privacy. However, the system faces a substantial hurdle, such as unexpectedly high gas consumption during PHE-related smart contract execution. The varying gas costs associated with encryption, additive, and decryption functions, with the additive operation incurring the highest cost, have emerged as a bottleneck, impacting both the economic feasibility and scalability of the BCFL system. These challenges underscore the pressing need to optimize gas consumption for PHE-related functions to ensure the continued effectiveness of the incentivization mechanism while maintaining privacy through HE.
   
\end{itemize}
\subsection{Future Directions}
The Federated Learning utilizing Blockchain technology offers numerous promising avenues for prospective research. We merely highlight specific directions for future investigations.
\begin{itemize}

    \item   \textbf{Privacy Preserving Techniques:} As a future research direction, explore the seamless integration of zero-knowledge proofs to bolster privacy within the Blockchain-enabled FL framework. This forward-looking methodology empowers participants to validate the accuracy of their updates without revealing the raw data, ensuring an elevated standard of confidentiality and privacy throughout the learning process. Moreover, it has the potential to substantially alleviate the verification burden on clients, thereby paving the way for more efficient and secure systems with enhanced privacy measures. Future research should focus on integrating zero-knowledge proofs in BCFL to enhance privacy, allowing participants to validate updates without revealing raw data. This approach not only ensures elevated confidentiality throughout the learning process but also has the potential to significantly reduce the verification burden on clients, paving the way for more efficient and secure systems with enhanced privacy measures.
    \item \textbf{Gas Consumption Optimization:} In the pursuit of advancing the BCFL system, a crucial area for exploration involves optimizing gas consumption for encryption-related smart contract functions. Currently, there is a notable absence of research addressing the reduction of gas consumption in the BCFL system, emphasizing privacy provision at a low cost. Investigating and implementing strategies to achieve low-cost and low gas consumption in the context of privacy-preserving BCFL operations represents a valuable and unexplored avenue for future research and system improvement.
    \item \textbf{Addressing Vulnerabilities in Smart Contracts:} For future directions in this area, it is crucial to prioritize research that delves explicitly into the vulnerabilities of smart contracts within BCFL. Given the inherent risks posed by faulty implementations, leading to persistent vulnerabilities and potential compromise of security and privacy in the model, a focused investigation is needed. Future efforts should aim to comprehensively identify and address these vulnerabilities, offering solutions to enhance the robustness of smart contracts in BCFL. This research would contribute to establishing a more secure foundation for the execution of logic and storage of final states in smart contracts, thereby mitigating risks associated with false data and bolstering overall security in BCFL models. Also, conduct a comprehensive security audit of the smart contract to identify and rectify system performance vulnerabilities.

\end{itemize}

\section{Summary and Conclusions}
\label{sec:conclusion}

Blockchain-enabled FL~(BCFL) systems are emerging approaches that combine the principles of FL with Blockchain technology to address various challenges. The main objective of these solutions is to guarantee privacy, security, and trust in decentralized collaborative Learning environments while providing a trustworthy and transparent framework for participants. By adopting a privacy perspective, this survey paper presents a systematic overview of the fundamental concepts of BCFL architectures and explores the opportunities and challenges that arise from their development. This survey gives a novel contribution to the present literature because it analyzes in detail the existing attacks on privacy in BCFL, along with state-of-the-art solutions relying on differential privacy, homomorphic encryption, secure multiparty computation, reward-driven approaches, multiple methods, and cross-chained FL.
Finally, we investigate the BCFL application in real-case scenarios, such as healthcare, Industry 5.0, and the Internet of Vehicles.
\begin{table}[ht]
\caption{Amount of papers analyzed per topic}
\centering
\begin{tabular}{|l|l|}
\hline
    \textbf{Topic} & \textbf{Amount of papers} \\
    \hline 
    BCFL Architecture & 16\\
    Attacks to privacy in BCFL & 2\\
    BCFL Architectures for Security and Privacy Protection & 14\\
    BCFL with Differential Privacy Approaches  & 19\\
    BCFL with HE Approaches  & 14\\
    BCFL with SMPC Approaches  & 5\\
    BCFL with Reward-driven Approaches  & 17\\
    BCFL with Hybrid Privacy Approaches  & 9\\
    Cross-chained FL Approaches for privacy  & 4\\
    Application of BCFL for privacy & 31\\
    \hline
\end{tabular}
\label{tab:summaryPerTopic}
\end{table}

\begin{figure}[ht]
	\centerline{
        \includegraphics[scale=0.5]{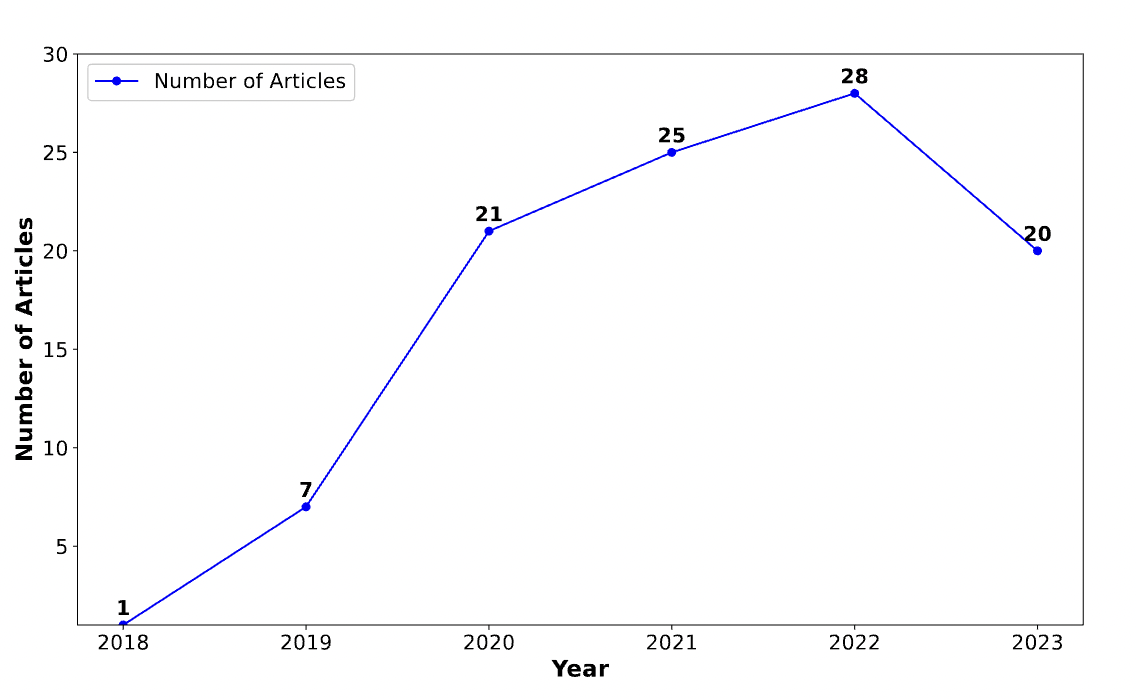}
    }
    \caption{Literature timeline}\label{fig:summaryPerYear}
\end{figure}

In summary, we analyzed 102 articles published in renowned international conferences, journals, symposiums, and workshops from 2018 to 2023 and focused on privacy aspects of Blockchain-enabled FL. Table \ref{tab:summaryPerTopic} represents a quantitative overview of the reviewed research papers divided into topics, whereas Figure~\ref{fig:summaryPerYear} pictures the analyzed number of articles published per year in the reference period.

The research direction explored in this paper can be regarded as a foundation, as we plan to continue our investigation by deep-diving into certain aspects only mentioned in the present work. For instance, a fascinating path can be the review of the paper exploiting existing security threats and countermeasures for BCFL systems to give the reader a larger spectrum of diverse problems in this domain. Moreover, an extensive and exhaustive technical description of all the implemented BCFL systems currently available is also a demanding task. 

We sincerely aspire for this work to assist researchers and practitioners in comprehending the essential aspects of this field, capturing notable advancements, and highlighting future research progress.

\section*{Acknowledgments}
This work was supported in part by the project SERICS (PE00000014) under the NRRP MUR program funded by the EU-NGEU, and by the Italian Ministry of University and Research through the PRIN Project ``HOMEY: a Human-centric IoE-based Framework for Supporting the Transition Towards Industry 5.0'' (code 2022NX7WKE), and by the HORIZON Europe Framework Programme through the project ``OPTIMA - Organization sPecific Threat Intelligence Mining and sharing'' (101063107).

%Bibliography
\bibliographystyle{unsrt}  
\bibliography{biblio}

\end{document}